\documentstyle[12pt]{article}
\begin{document}
\thispagestyle{empty}
\newcommand{\be}{\begin{equation}}
\newcommand{\ee}{\end{equation}}
\newcommand{\sect}[1]{\setcounter{equation}{0}\section{#1}}
\newcommand{\vs}[1]{\rule[- #1 mm]{0mm}{#1 mm}}
\newcommand{\hs}[1]{\hspace{#1mm}}
\newcommand{\mb}[1]{\hs{5}\mbox{#1}\hs{5}}
\newcommand{\bea}{\begin{eqnarray}}
\newcommand{\eea}{\end{eqnarray}}
\newcommand{\wt}[1]{\widetilde{#1}}
\newcommand{\ux}[1]{\underline{#1}}
\newcommand{\ov}[1]{\overline{#1}}
\newcommand{\sm}[2]{\frac{\mbox{\footnotesize #1}\vs{-2}}
           {\vs{-2}\mbox{\footnotesize #2}}}
\newcommand{\prt}{\partial}
\newcommand{\eps}{\epsilon}\newcommand{\p}[1]{(\ref{#1})}
\newcommand{\R}{\mbox{\rule{0.2mm}{2.8mm}\hspace{-1.5mm} R}}
\newcommand{\Z}{Z\hspace{-2mm}Z}
\newcommand{\cd}{{\cal D}}
\newcommand{\cg}{{\cal G}}
\newcommand{\ck}{{\cal K}}
\newcommand{\cw}{{\cal W}}
\newcommand{\vj}{\vec{J}}
\newcommand{\vl}{\vec{\lambda}}
\newcommand{\vz}{\vec{\sigma}}
\newcommand{\vt}{\vec{\tau}}
\newcommand{\poiss}{\stackrel{\otimes}{,}}
\newcommand{\tx}{\theta_{12}}
\newcommand{\tb}{\overline{\theta}_{12}}
\newcommand{\zw}{{1\over z_{12}}}
\newcommand{\sqp}{{(1 + i\sqrt{3})\over 2}}
\newcommand{\sqm}{{(1 - i\sqrt{3})\over 2}}
\newcommand{\NP}[1]{Nucl.\ Phys.\ {\bf #1}}
\newcommand{\PLB}[1]{Phys.\ Lett.\ {B \bf #1}}
\newcommand{\PLA}[1]{Phys.\ Lett.\ {A \bf #1}}
\newcommand{\NC}[1]{Nuovo Cimento {\bf #1}}
\newcommand{\CMP}[1]{Commun.\ Math.\ Phys.\ {\bf #1}}
\newcommand{\PR}[1]{Phys.\ Rev.\ {\bf #1}}
\newcommand{\PRL}[1]{Phys.\ Rev.\ Lett.\ {\bf #1}}
\newcommand{\MPL}[1]{Mod.\ Phys.\ Lett.\ {\bf #1}}
\newcommand{\BLMS}[1]{Bull.\ London Math.\ Soc.\ {\bf #1}}
\newcommand{\IJMP}[1]{Int.\ J.\ Mod.\ Phys.\ {\bf #1}}
\newcommand{\JMP}[1]{Jour.\ Math.\ Phys.\ {\bf #1}}
\newcommand{\LMP}[1]{Lett.\ Math.\ Phys.\ {\bf #1}}
\renewcommand{\thefootnote}{\fnsymbol{footnote}}
\newpage
\setcounter{page}{0}
\pagestyle{empty}
\vs{12}
\begin{center}
{\LARGE {\bf The Signature Triality of}}\\ {\LARGE {\bf
Majorana-Weyl Spacetimes}}\\ [0.8cm]

\vs{10} {\large M.A. De Andrade$^{(a,b)}$, M. Rojas$^{(a)}$ and F.
Toppan$^{(a)}$} ~\\ \quad \\
 {\em {~$^{(a)}$} CBPF, DCP, Rua Dr. Xavier Sigaud
150, cep 22290-180 Rio de Janeiro (RJ), Brazil}\\
~\quad \\
 {\em {~$~^{(b)}$} UCP, FT, Rua
Bar{\~{a}}o do Amazonas, 124, cep 25685-070, Petr\'{o}polis (RJ),
Brazil}
\end{center}
\vs{6}

\centerline{ {\bf Abstract}}

\vs{6}

Higher dimensional Majorana-Weyl spacetimes present space-time
dualities which are induced by the $Spin(8)$ triality
automorphisms. Different signature versions of theories such as
$10$-dimensional SYM's, superstrings, five-branes, F-theory, are
shown to be interconnected via the $S_3$ permutation group.
Bilinear and trilinear invariants under space-time triality are
introduced and their possible relevance in building models
possessing a space-versus-time exchange symmetry is discussed.
Moreover the Cartan's ``vector/chiral spinor/antichiral spinor"
triality of $SO(8)$ and $SO(4,4)$ is analyzed in detail and
explicit formulas are produced in a Majorana-Weyl basis. This
paper is the extended version of hep-th/9907148. \vs{6} \vfill
\rightline{CBPF-NF-009/00}
\rightline{hep-th/xxx} {\em E-Mails:\\ 1) marco@cbpf.br\\ 2)
mrojas@cbpf.br \\ 3) toppan@cbpf.br}
\newpage
\pagestyle{plain}
\renewcommand{\thefootnote}{\arabic{footnote}}
\setcounter{footnote}{0}
\vs{8}

\section{Introduction.}

Physical theories formulated in different-than-usual spacetimes
signatures have recently found increased attention. One of the
reasons can be traced to the conjectured $F$-theory \cite{Vafa}
which supposedly lives in $(2+10)$ dimensions \cite{Nish}. The
current interest in AdS theories motivated by the AdS/CFT
correspondence furnishes another motivation. Two-time physics e.g.
has started been explored by Bars and collaborators in a series of
papers \cite{Bars}. From another point of view we can also recall
that a fundamental theory is expected to explain not only the
spacetime dimensionality, but even its signature (see
\cite{Duff}). Quite recently Hull and Hull-Khuri \cite{Hull}
pointed out the existence of dualities relating different
compactifications of theories formulated in different signatures.
Such a result provides new insights to the whole question of
spacetime signatures. In another context (see e.g. \cite{Cori})
the existence of space-time dualities has also been remarked.\par
Majorana-Weyl spacetimes (i.e. those supporting Majorana-Weyl
spinors) are at the very core of the present knowledge of the
unification via supersymmetry, being at the basis of
ten-dimensional superstrings, superYang-Mills and supergravity
theories (and perhaps the already mentioned $F$-theory). A
well-established feature of Majorana-Weyl spacetimes is that they
are endorsed of a rich structure. A legitimate question that could
be addressed is whether they are affected, and how, by space-time
dualities. The answer is positive. Indeed all different
Majorana-Weyl spacetimes which are possibly present in any given
dimension are each-other related by duality transformations which
are induced by the $Spin(8)$ triality automorphisms. The action of
the triality automorphisms is quite non-trivial and has far richer
consequences than the ${\bf Z}_2$-duality (its most trivial
representative) associated to the space-time $(s,t)\leftrightarrow
(t,s)$ exchange discussed in \cite{Duff}. It corresponds to $S_3$,
the six-element group of permutations of three letters, identified
with the group of congruences of the triangle and generated by two
reflections. The lowest dimension in which the triality action is
non-trivial is $8$ (not quite a coincidence), where the spacetimes
$(8+0)-(4+4)-(0+8)$ are all interrelated. They correspond to the
transverse coordinates of the $(9+1)-(5+5)-(1+9)$ spacetimes
respectively, where the triality action can also be lifted.
Triality relates as well the $12$-dimensional Majorana-Weyl
spacetimes $(10+2)-(6+6)-(2+10)$, i.e. the potentially interesting
cases for the $F$-theory, and so on. Triality allows explaining
the presence of points (read theory) in the brane-molecule table
of ref. \cite{Duff}, corresponding to the different versions of
e.g. superstrings, $11$-dimensional supermembranes, fivebranes.
\\ As a consequence of triality, supersymmetric theories
formulated with Majorana-Weyl spinors in a given dimension but
with different signatures, are all dually mapped one into another.
A three-language dictionary is here furnished with the exact
translations among the different versions of such supersymmetric
theories. It should be stressed that, unlike \cite{Hull}, the
dualities here discussed are already present for the {\em
uncompactified} theories and in this respect look more
fundamental.\par The reason why the triality of the $d=8$
dimension plays a role in dually relating higher dimensional
spacetimes is a consequence of the fact that higher-dimensional
Lorentz groups admit the $8$-dimensional Lorentz as a subgroup.
This feature is neatly encoded in the representation properties of
the higher dimensional Clifford $\Gamma$-matrices. In fact, due to
this argument, it can be shown that not only different-signature
Majorana-Weyl spacetimes are duality-related, but also
that different-signatures odd-dimensional Majorana spacetimes are all
interconnected via triality. This is true in particular for the
$11$-dimensional case relevant for the maximal supergravity and
the $M$-theory.\par Manifestations of triality are observed in
different contexts. We will stress the fact that besides the
original ``Cartan" triality \cite{Cart} exhibited by vectors,
chiral and antichiral spinors (and therefore also denoted as
``VCA"-triality in the following) in each one of the three
Majorana-Weyl spacetimes of signature $(8+0)$, $(4+4)$, $(0+8)$
respectively, consequences of triality are found at the Clifford's
$\Gamma$ matrices representation level. In $d=8$ dimension this is
seen by the fact that ``Majorana type representations" for
$\Gamma$-matrices, i.e. such that all the $\Gamma$'s have a
definite (anti-)symmetry property, only exist for the $4_S+4_A$,
$8_S+0_A$, $0_S+8_A$ cases.\par Moreover it can be shown that the
three Majorana-Weyl spacetimes of signatures $(4+4)$, $(8+0)$,
$(0+8)$ are interrelated via the $S_3$ permutation group. We call
this property the ``signature-triality" or the ``space-time
triality".\par It is worth stressing the fact that the arising of
the $S_3$ permutation group as a signature-duality group for
Majorana-Weyl spacetimes in a given dimension {\em is not} a
completely straightforward consequence of the existence of
Majorana-Weyl spacetimes in three different signatures. Some
extra-requirements have to be fulfilled in order this to be true.
As an example we just mention that a necessary condition for the
presence of $S_3$ requires that each given couple of the three
different spacetimes must differ by an {\em even} number of
signatures (in the text this point will be discussed in full
detail); the flipping of an {\em odd} number of signatures, like
in the Wick rotation from Minkowski to the Euclidean space, cannot
be achieved with a ${\bf Z}_2$ group when spinors are involved.
\par
In the present paper various aspects of the triality property will
be rather extensively discussed. This is due to the objective
relevance of the transverse $8$-dimensional space of coordinates
in the light-cone formulations of superstrings theories, branes,
and the non-perturbative (M)atrix approach to $M$-theory. The
eight dimensions are indeed the natural setting for the appearance
of triality-related symmetries. Triality is of course a very
well-known property which has been extensively investigated in the
literature both for technical reasons (as an example it implies
the equivalence of the NSR and GS formulation of superstrings, see
e.g. \cite{Nepo}) as well in its more fundamental aspects. In this
paper, besides discussing the signature triality, we express its
action on the Majorana-type representations of the Clifford's
$\Gamma$ matrices. Moreover, for completeness, we review and
extend some of the original Cartan's results on V-C-A triality.

He gave a concrete representation within a non-diagonal metric
which, once diagonalized, shows the $(4+4)$-signature. We point
out that the triality generators he produced {\em do not} close
the $S_3$ permutation group. However, an alternative presentation
is available which implies the existence of $S_3$. As a
consequence the triality group ${\cal G}_{Tr}$ defined in section
$6$ is given by the semidirect product of a linear subgroup ${\cal
G}$ (investigated also in \cite{Gamba}) of $24$-dimensional
matrices with such an $S_3$ permutation group. It is worth
mentioning that in the formulas of paragraph 139 of Cartan's book
an obvious typographical mistake plus a sign error appear.
\par
In the
appendices $2$-$4$ we produce the triality generators closing the
$S_3$ group for each one of the three signatures $(4+4)$, $(8+0)$,
$(0+8)$ which carry the triality structure. All formulas are here
expressed in a Majorana-Weyl basis.
\par
Furthermore, the construction of bilinear and trilinear invariants
under the $S_3$ permutation group of the three Majorana-Weyl
spacetimes is performed. They can be possibly used to formulate
supersymmetric Majorana-Weyl theories in a manifestly
triality-invariant form which presents an explicit symmetry under
exchange of space and time coordinates.\par The scheme of this
work is as follows. In the next section we recall, following
\cite{Kugo} and \cite{DeAn}, the basic properties of
$\Gamma$-matrices and Majorana conditions needed for our
construction. Majorana-type representations are analyzed in
section $3$. We show there how to relate the Majorana-Weyl
representations in $d> 8$ to the $8$-dimensional Majorana-Weyl
representations. In section $4$ we introduce, for $d=8$, the set
of data necessary to define a supersymmetric Majorana-Weyl theory,
i.e. the set of ``words" of our three-languages dictionary. The
Cartan's \cite{Cart} triality among vectors, chiral and antichiral
spinors is presented in section $5$. The main result is furnished
in section $6$, where spacetime triality is discussed. In section
$7$ the triality as a generator of a group of invariances is
analyzed. In section $8$ an application of triality is made. It is
shown how to connect  via triality some points (theories)
presented in the Blencowe-Duff ``brane-molecule scan" of reference
\cite{Duff} (further enlarged in the second paper referred in
\cite{Duff}). In the original paper only the presence of ``mirror"
theories connected by a ${\bf Z}_2$ space-versus-time exchange was
``explained". In the Conclusions we furnish some comments and
point out some perspectives of future works. In the appendices,
besides the already mentioned results, some useful construction
concerning $\Gamma$ matrices representations in $6$ and $8$
dimensions is given.

\vspace{0.2cm} \noindent{\section{Preliminary results.}}

In this section the basic ingredients
needed
for our construction and the conventions employed
will be introduced.
More detailed information can be found in
\cite{Kugo}
and \cite{DeAn}.\par We denote as $g_{mn}$ the flat
(pseudo-)euclidean metric of a $(t+s)$-spacetime. Time (space)
directions in our conventions are associated to the $+$
(respectively $-$) sign. \par The $\Gamma$'s matrices are assumed
to be unitary and without loss of generality a time-like $\Gamma$ is
normalized so that its square is $+1$. The three matrices ${\cal
A}$, ${\cal B}$, ${\cal C}$ are the generators of the three
conjugation operations (hermitian, complex conjugation and
transposition respectively) on the $\Gamma$'s. In particular
\begin{eqnarray}
{\cal C} \Gamma^m {\cal C}^\dagger &=& \eta(-1)^{t+1}{\Gamma^m}^T
\label{0}
\end{eqnarray}
where $\eta=\pm 1$ is a sign. In even-dimensional spacetimes it
labels the two inequivalent choices of the charge conjugation
matrix ${\cal C}$.\par A relation exists, given by the formula
\begin{eqnarray}
{\cal C}&=& {\cal B}^T {\cal A} \label{1}
\end{eqnarray}
expressing anyone of the three matrices ${\cal A}, {\cal B}, {\cal
C}$ in terms of the two others.\par Up to an inessential phase,
${\cal A}$ is specified by the product of all the time-like
$\Gamma$ matrices.\par An unitary transformations $U$ applied on
spinors act on $\Gamma^{m}$, ${\cal A}, {\cal B}, {\cal C}$
according to \cite{DeAn}
\begin{eqnarray}
\Gamma^m &\mapsto& U\Gamma^m U^\dag\nonumber\\ {\cal A}&\mapsto&
U{\cal A}U^{\dag}\nonumber\\{\cal B}&\mapsto& U^\ast {\cal B}
U^{\dag}\nonumber\\
{\cal C}&\mapsto & U^\ast{\cal C} U^\dag \label{2}
\end{eqnarray}
A Majorana representation for the $\Gamma$'s can be defined as the one
in which ${\cal B}$ is set equal to the identity. Spinors can be
assumed real in this case.\par
In even dimensions we can also introduce the notion of Weyl
representation, i.e. when the ``generalized $\Gamma^5$ matrix" is
symmetric and block diagonal and with no loss of generality
can be assumed to be the direct sum of the two equal-size blocks
${\bf 1}\oplus (-{\bf 1})$.
The compatibility of both Majorana and Weyl conditions
constraints the spacetime $(t+s)$ to satisfy
\begin{eqnarray}
s-t&=& 0 \quad {\it mod}\quad 8, \quad {\it for} \quad{\it both}\quad
{\it values}
\quad \eta=\pm 1
\label{3}
\end{eqnarray}
In even dimensions
Majorana representations, but not of Weyl type, are also found for
\begin{eqnarray}
s-t &=& 2 \quad  mod \quad 8 \quad  for \quad \eta=-1;\nonumber \\
s-t &=& 6 \quad mod \quad 8 \quad for \quad \eta= +1.
\label{4}
\end{eqnarray}
In odd dimensions Weyl spinors cannot be defined, while Majorana
spinors exist for \begin{eqnarray} s- t &=& 1,7 \quad mod \quad 8
\label{5}
\end{eqnarray}
For $d<8$ the only spacetimes supporting Majorana-Weyl spinors
have signatures $(n+n)$. At $d=8$ a new feature arises,
Majorana-Weyl spinors can be found for different signatures.
\par
Making explicit the relation between theories formulated in
such different signatures is the main content of this paper.
{\quad}\\

 \vspace{0.2cm}
\noindent{\section{Majorana-type representations.}}

It is convenient to introduce the notion of Majorana-type
representation (or shortly MTR) of the Clifford' s $\Gamma$
matrices. It can be defined as a representation such that all the
$\Gamma$'s have a definite symmetry, being either symmetric or
antisymmetric. In $d$ dimensions a MTR with $p$ symmetric and $q$
antisymmetric $\Gamma$'s ($p+q=d$) will be denoted as $(p_S +
q_A)$ in the following. \par When specialized to such
representations the ${\cal C}$ charge-conjugation matrix
introduced in the previous section is given by either the product
of all the symmetric $\Gamma$ matrices, denoted as ${\cal C}_S$,
or all the antisymmetric ones (${\cal C}_A$)
\begin{eqnarray}
{\cal C}_S &=& \Pi_{i=1,..., p}{\Gamma^i}_S\nonumber\\
{\cal C}_A &=& \Pi_{i=1,...,q} {\Gamma^i}_A
\end{eqnarray}
Please notice that the index $S,A$ is not referred to the
(anti-)symmetry property of the matrices $C_{S,A}$ themselves.\par
In even dimensions ${\cal C}_S$, ${\cal C}_A$ correspond to
opposite values of $\eta$ in (\ref{0}), while in odd dimensions,
up to an inessential phase factor, the two definitions for the
charge-conjugation matrix collapse into a single matrix. This is
in agreement with the property that in odd dimensions, up to
unitary conjugation, the ${\cal C}$-matrix is uniquely defined.
The convenience of using MTR's to discuss Wick rotations to and
from the Euclidean space has been advocated in \cite{DeAn}.\par It
can be easily recognized that a Majorana representation for
Clifford's $\Gamma$ matrices in a given signature spacetime
implies the $\Gamma$'s belong to a MTR. Conversely, given a MTR
with $p_S, q_A$ (anti-)symmetric $\Gamma$ matrices, two spacetimes
exist ($t=p$, $s=q$ with the choice ${\cal C}\equiv {\cal C}_S$,
and respectively $t=q$, $s=p$ for ${\cal C}\equiv {\cal C}_A$)
such that the representation is Majorana (i.e. ${\cal B}={\bf
1}$). The admissible couples of $(p_S, q_A)$ values for a MTR can
be immediately read from the Majorana tables given above
({\ref{3}), (\ref{4}) and (\ref{5}). The construction is such that
${\cal C}$ must correspond to the correct value of $\eta$
appearing in the tables.\par The list of all possible MTR's in any
given dimension is therefore easily computed. In order to furnish
an example we mention that in $d=6$ there exists a MTR (not of
Weyl kind) with $6$ anticommuting $\Gamma$ matrices plus an
anticommuting $\Gamma^7$ $(0_S + 6_A, {\Gamma^7}_A)$. It provides
a Majorana basis for an Euclidean $6$-dimensional space. A
concrete realization of such a representation is presented in
appendix $1$\footnote{It is worth mentioning that such a
representation can be written in terms of the octonionic structure
constants (see ref. \cite{gk}). In the present work our results
have been obtained making no explicit reference to octonions;
investigating the connection with such a division algebra is
outside the scope of this work.}.\par For completeness we present
a list of all MTR's up to $d=18$. We obtain
\begin{eqnarray}
&&
\begin{array}{cccc}
{\bf d} & {\bf W\cdot R} & &{\bf NW\cdot R} \\ \hline 2+1 &
(1_S+1_A+{\Gamma^3}_S)&\leftrightarrow & (
  2_S+0_A+{\Gamma^3}_A) \\ \hline
   4+1 & (2_S+2_A+{\Gamma^5}_S)& \leftrightarrow
  & (3_S+1_A+{\Gamma^5}_A) \\ \hline
  6+1 & (3_S+3_A+{\Gamma^7}_S) & \leftrightarrow &(4_S+2_A+{\Gamma^7}_A)
  \\
  6+1 &  &  & (0_S+6_A+{\Gamma^7}_A) \\ \hline
  8+1 &(8_S+0_A+{\Gamma^9}_S)  &  &  \\
  8+1 &(4_S+4_A+{\Gamma^9}_S)  & \leftrightarrow &(5_S+3_A+{\Gamma^9}_A)  \\
  8+1 &(0_S+8_A+{\Gamma^9}_S)  & \leftrightarrow & (1_S+7_A+{\Gamma^9}_A) \\
\hline
  10+1 &(9_S+1_A+{\Gamma^{11}}_S)  & \leftrightarrow &
  (10_S+0_A+{\Gamma^{11}}_A)  \\
  10+1 &(5_S+5_A+{\Gamma^{11}}_S)  & \leftrightarrow &
  (6_S+4_A+{\Gamma^{11}}_A)  \\
  10+1 &(1_S+9_A+{\Gamma^{11}}_S)  & \leftrightarrow &
  (2_S+8_A+{\Gamma^{11}}_A)  \\ \hline
  12+1 &({10}_S+2_A+{\Gamma^{13}}_S)  & \leftrightarrow &
  ({11}_S+1_A+{\Gamma^{13}}_A)  \\
  12+1 & ({6}_S+6_A+{\Gamma^{13}}_S) & \leftrightarrow &
  ({7}_S+5_A+{\Gamma^{13}}_A) \\
  12+1 &({2}_S+{10}_A+{\Gamma^{13}}_S)  & \leftrightarrow &
  ({3}_S+{9}_A+{\Gamma^{13}}_A)  \\ \hline
  14+1 &({11}_S+3_A+{\Gamma^{15}}_S)  & \leftrightarrow &
  ({12}_S+2_A+{\Gamma^{15}}_A)  \\
  14+1 & ({7}_S+7_A+{\Gamma^{15}}_S) & \leftrightarrow &
  ({8}_S+6_A+{\Gamma^{15}}_A) \\
  14+1 & ({3}_S+{11}_A+{\Gamma^{15}}_S) & \leftrightarrow &
  ({4}_S+{10}_A+{\Gamma^{15}}_A)  \\
  14+1 &   &                 &({0}_S+{14}_A+{\Gamma^{15}}_A)   \\ \hline
  16+1 &({16}_S+0_A+{\Gamma^{17}}_S)  &  &  \\
  16+1 & ({12}_S+4_A+{\Gamma^{17}}_S) & \leftrightarrow &
  ({13}_S+3_A+{\Gamma^{17}}_A)  \\
  16+1 & ({8}_S+8_A+{\Gamma^{17}}_S) & \leftrightarrow &
  ({9}_S+7_A+{\Gamma^{17}}_A)  \\
  16+1 & ({4}_S+{12}_A+{\Gamma^{17}}_S) & \leftrightarrow &
  ({5}_S+{11}_A+{\Gamma^{17}}_A)  \\
  16+1 & ({0}_S+{16}_A+{\Gamma^{17}}_S) & \leftrightarrow &
  ({1}_S+{15}_A+{\Gamma^{17}}_A)  \\ \hline
  18+1 &({17}_S+1_A+{\Gamma^{19}}_S)  & \leftrightarrow &
  ({18}_S+0_A+{\Gamma^{19}}_A)  \\
  18+1 & ({13}_S+5_A+{\Gamma^{19}}_S) & \leftrightarrow &
  ({14}_S+4_A+{\Gamma^{19}}_A)  \\
  18+1 & ({9}_S+9_A+{\Gamma^{19}}_S) & \leftrightarrow &
  ({10}_S+8_A+{\Gamma^{19}}_A)  \\
  18+1 & ({5}_S+{13}_A+{\Gamma^{19}}_S) & \leftrightarrow &
  ({6}_S+{12}_A+{\Gamma^{19}}_A)  \\
  18+1 & ({1}_S+{17}_A+{\Gamma^{19}}_S) & \leftrightarrow &
  ({2}_S+{16}_A+{\Gamma^{19}}_A) \\ \hline
       & & &
\end{array}
\end{eqnarray}
Some comments are in order. In the second column we listed all
MTR's of Weyl type, in the last one the non-Weyl representations.
An arrow connects two given Weyl and non-Weyl representations
which are intertwined by an exchange of the ``generalized
$\Gamma^5$" matrix with any other $\Gamma$-matrix of opposite
symmetry. Up to $d=18$ the only true genuine even-dimensional
non-Weyl Majorana representations not having such a Weyl
counterpart are the above-mentioned $6$-dimensional $(0_S+6_A)$
representation and the $14$-dimensional $(0_S+14_A)$, both
appearing in the table. Their difference in dimensionality $(=8)$
is of course a consequence of the famous $mod\quad 8$ property of
the $\Gamma$ matrices, see e.g. \cite{Choq}.\par Different MTR's
belong to different classes under similarity transformations of
the $\Gamma$'s representations. Indeed in, let's say, an euclidean
(all $+$ signs) space, the index
\begin{eqnarray}
&&I = tr (\Gamma^m \cdot {\Gamma_m}^T) = (p_S-q_A)\cdot tr{\bf 1}
\label{index}
\end{eqnarray}
takes different values for different MTR's and is by construction
invariant under the transformation $\Gamma\mapsto O \Gamma O^T$
realized by orthogonal matrices $O$.\par Up to $d=8$ (excluded)
there exists a unique similarity class of MTR's of Weyl type, so
that Majorana-Weyl spinors can be defined in $(n+n)$ space-times
only. A new feature arises starting from $d\geq 8$, Weyl and
Majorana-type representations are compatible for different
similarity classes. In $d=8$ the three solutions $(8_S+0_A)$,
$(4_S+4_A)$, $(0_S+8_A)$, associated with the corresponding
Majorana-Weyl spacetimes, are found.
\par
An efficient tool to produce and analyze Weyl representations in
any higher dimensions is furnished by the following algorithm
\cite{Cola} and \cite{DeAn}. It provides a recursive procedure to
construct $d$-dimensional Weyl representations in terms of any
given couple of $r$, $s$ lower-dimensional $\Gamma$-matrices
representations, where the even integers $d,r,s$ are constrained
to satisfy
\begin{eqnarray}
d &=& r+s +2
\end{eqnarray}
Moreover, if the $r,s$-dimensional representations are of
Majorana-type, the $d$-dimensional one is Majorana-Weyl. \par The
algorithm can be expressed through the formula
\begin{eqnarray}
{\Gamma_d}^{i=1,..., s+1} &=& \sigma_x\otimes {\bf 1}_L\otimes
{\Gamma_s}^{i=1,...,s+1}\nonumber\\ {\Gamma_d}^{s+1+j = s+2, ...,
d} &=& \sigma_y\otimes{\Gamma_r}^{j=1,...,r+1}\otimes {\bf 1}_R
\label{algo}
\end{eqnarray}
where ${\bf 1}_{L,R}$ are the unit-matrices in the respective
spaces, while $\sigma_x = e_{12}+e_{21}$ and $\sigma_y = -i
e_{12}+i e_{21}$ are the $2$-dimensional Pauli matrices.
${\Gamma_r}^{r+1}$ corresponds to the ``generalized
$\Gamma^5$-matrix" in $r+1$ dimensions. In the above formula the
values $r,s=0$ are allowed. The corresponding ${\Gamma_0}^1$ is
just $1$.\par By iteratively applying the above algorithm
starting from $1$ (that
is $r,s=0$), we obtain as a first step the
$2$-dimensional
Pauli matrices and next, from $r=0$, $s=2$, the
$4$-dimensional MW representation as a second step.
The $6$-dimensional MW
representation ($3_S+3_A$) is obtained as a further step.
It can be produced either from $r=0$, $s=4$, or $r=2$, $s=2$. The
non-Weyl ($0_S+6_A$) representation is constructed with the
method explained in appendix $1$.\par
The higher-dimensional MW
representations as well are obtained via the algorithm.
An explicit way of constructing them is, e.g., in accordance
with the table
\begin{eqnarray}
\begin{array}{ccccccc}
r=6 & (0_S+6_A) &\quad  & s=0 &(1) & \mapsto & (8_S+0_A) \\ r =6 &
(3_S+3_A)&\quad & s=0 &(1) & \mapsto & (4_S+4_A)  \\ r=0 & (1)
&\quad & s=0 & (0_S+6_A) & \mapsto & (0_S+8_A) \\ \hline r=0 & (1)
&\quad & s=8 & (8_S+0_A) & \mapsto & (9_S+1_A)
\\ r=8 &(4_S+4_A)&\quad & s=0 & (1) & \mapsto & (5_S+5_A) \\
r=8 & (8_S+0_A)&\quad & s=0 & (1) & \mapsto & (1_S+9_A) \\ \hline
r=0 & (1)&\quad & s=10 & (9_S+1_A) & \mapsto & (10_S+2_A) \\ r=0 &
(1)&\quad & s=10 & (5_S+5_A) & \mapsto & (6_S+6_A) \\ r=0 &
(1)&\quad & s=10 & (1_S+9_A) & \mapsto & (2_S+10_A) \\ \hline r=0
& (1)&\quad & s=12 & (10_S+2_A) & \mapsto & (11_S+3_A) \\ r=0 &
(1)&\quad & s=12 & (6_S+6_A) & \mapsto & (7_S+7_A) \\ r=0 &
(1)&\quad & s=12 & (2_S+10_A) & \mapsto & (3_S+11_A) \\ \hline
r=6 & (0_S +6_A)&\quad & s=8 & (8_S+0_A) & \mapsto & (16_S+0_A)\\
r=0 & (1)&\quad & s=14 & (11_S+3_A) & \mapsto & (12_S+4_A) \\ r=0
& (1)&\quad & s=14 & (7_S+7_A) & \mapsto & (8_S+8_A) \\ r=0 &
(1)&\quad & s=14 & (3_S+11_A) & \mapsto & (4_S+12_A) \\ r=8 &
(8_S+0_A)&\quad & s=6 & (0_S+6_A) & \mapsto & (0_S+16_A)\\
\hline r=0 & (1)&\quad & s=16 & (16_S+0_A) & \mapsto &
(17_S+1_A)\\ r=0 & (1)&\quad & s=16 & (12_S+4_A) & \mapsto &
(13_S+5_A)\\
 r=0 & (1)&\quad & s=16 & (8_S+8_A) & \mapsto &
(9_S+9_A)\\
 r=0 & (1)&\quad & s=16 & (4_S+12_A) & \mapsto &
(5_S+13_A)\\
 r=0 & (1)&\quad & s=16 & (0_S+16_A) & \mapsto &
(1_S+17_A)
\end{array}
&&
\end{eqnarray}
\par
Notice that in order to explicitly construct all
MW-representations up to $d=18$ with the help of formula
(\ref{algo}), the only extra-knowledge of the $(0_S+6_A$) non-Weyl
Majorana is required.
\par
In the following we will refer to ``space-time triality" as the
$8$-dimensional property that the Majorana-Weyl condition is
satisfied in three different signatures and will show its
connection to the usual triality property of the $8$ dimensions,
as well as the $S_3$ permutation group.\par Due to the ``lifting"
formula (\ref{algo}), this feature is extended to
higher-dimensional MW-spacetimes. Solutions in three different
signatures arise as well in $d=10$, $d=12$ and $d=14$. Such
solutions are a direct consequence of the embedding of the
eight-dimensional Lorentz algebra into higher dimensions. In
dimensions higher than $8$ the property that Majorana-Weyl
conditions (or simply Majorana conditions in odd dimensions) are
consistent in different signatures can therefore be regarded as a
``derived" property which is fundamentally rooted in the
$8$-dimensions. This argument holds even in the case (for $d\geq
16$), where solutions to the MW-constraints in more than three
different signatures are obtained. As an example the $d=18$ case
can be produced with the help of (\ref{algo}) for the values
$r=s=8$. Therefore the $5$ different $18$-dimensional
Majorana-Weyl representations are obtained from tensoring two
$8$-dimensional Majorana-Weyl representations.\par On a physical
ground and not just for purely mathematical purposes, at the
present state of the art we do not need getting involved into such
complications since the most promising dimensions where the
ultimate candidate theories for unification are expected to live
correspond to $d=10,11,12$.\par As a final comment in this section
we emphasize once more that all data needed to define theories in
such dimensions can be recovered, through a set of reconstruction
formulas based on (\ref{algo}), from the $8$-dimensional data. In
particular all ``space-time trialities" in $d>8$ are encoded in
the $8$-dimensional ``space-time triality". For this reason in the
following we can concentrate ourselves in investigating the
$8$-dimensional case.

\vspace{0.2cm} \noindent{\section{The set of data for
Majorana-Weyl supersymmetric theories.}}

In this section we present the set of data needed to specify a
supersymmetric theory involving Majorana-Weyl spinors. The most
suitable basis one can use in this case is the Majorana-Weyl basis
previously discussed, where all spinors are either real or
imaginary. In such a representation the following set of data
underlines any given theory:\\ {\em{\bf i)}} the vector fields
(or, in the string/brane picture, the bosonic coordinates of the
target $x_m$), specified by a vector index denoted by $m$;\\
{\em{\bf ii)}} the spinor fields (or, in the string/brane picture,
the fermionic coordinates of the target $\psi_a$,
$\chi_{\dot{a}}$), specified by chiral and antichiral indices $a$,
$\dot{a}$ respectively;\\ {\em{\bf iii)}} the diagonal
(pseudo-)orthogonal spacetime metric $(g^{-1})^{mn}$, $g_{mn}$
which we will assume to be flat;\\ {\em{\bf iv)}} the ${\cal A}$
matrix introduced in section $2$, used to define barred spinors,
coinciding with the $\Gamma^0$-matrix in the Minkowski case; in a
MWR is decomposed in an equal-size block diagonal form such as
${\cal A} = A\oplus {\tilde A}$, with structure of indices
${(A)_a}^b$ and ${(\tilde{A})_{\dot{a}}}^{\dot{b}}$
respectively;\\ {\em{\bf v)}} the charge-conjugation matrix ${\cal
C}$ which also appears in an equal-size block diagonal form ${\cal
C} = {C^{-1}}\oplus {\tilde C}^{-1}$. It is invariant under
bispinorial transformations and it can be promoted to be a metric
in the space of chiral (and respectively antichiral) spinors, used
to raise and lower spinorial indices. Indeed we can set
$(C^{-1})^{ab}$, $(C)_{ab}$, and $({\tilde
C}^{-1})^{\dot{a}\dot{b}}$, $({\tilde C})_{\dot{a}\dot{b}}$;\\
{\em {\bf vi)} } the $\Gamma$-matrices, which are decomposed in
equal-size blocks as in (\ref{blocks}), where the $\sigma^m$'s are
upper-right blocks and the ${\tilde \sigma}^m$'s lower-left blocks
having structure of indices ${(\sigma^m)_a}^{\dot{b}}$ and
${(\tilde{\sigma}^m)_{\dot{a}}}^b$ respectively;\\ {\em{\bf vii)}}
the $\eta=\pm 1$ sign, labeling the two inequivalent choices for
${\cal C}$.\par We recall that by definition the ${\cal B}$ matrix
is automatically set to be the identity (${\cal B} ={\bf 1}$) in a
Majorana-Weyl representation.\par The above structures are common
in any theory involving Majorana-Weyl spinors. In the following we
will furnish a dictionary relating Majorana-Weyl spacetimes with
the same dimensionality, but with different signatures. The
structures {\em\bf i)}}-{\em{\bf vii)}} will be related via
triality transformations which close the $S_3$ permutation group.
They constitute the ``words" in a three-language dictionary.
According to the discussion in the previous section without loss
of generality we can limit ourselves in analyzing the
$d=8$-dimensional case. In this particular dimension the three
indices, vector ($m$), chiral ($a$) and antichiral ($\dot a$) take
values $m,a , {\dot a} \in \{ 1,...,8\}$.\par We mention that in
the $(4+4)$-signature the $(4_S+4_A)$-representation of the
$\Gamma$-matrices has to be employed for both values of $\eta$ in
order to provide a Majorana-Weyl basis. In the ($t=8$, $s=0$)
signature the $(8_S+0_A)$-representation offers a MW basis for
$\eta =+1$, while the $(0_S+8_A)$ offers it for $\eta = -1$. The
converse is true in the ($t=0$, $s=8$)-signature.\par The three
$8$-dimensional MW-type representations are explicitly constructed
in the appendices $2$-$4$, together with their charge-conjugation
matrices ${\cal C}$'s. \vspace{0.2cm} \noindent{\section{The
Cartan's V-C-A triality.}}

In this section we review the basic features of the Cartan's
triality involving vectors, chiral and antichiral spinors of
$SO(8)$ and $SO(4,4)$.\par The fundamental reason behind triality
is the peculiar property of the $D_4$ Lie algebra, the only one
admitting a group of symmetry for the corresponding Dynkin diagram
other than the identity or ${\bf Z}_2$. Its group of symmetry is
the six-elements non-abelian group of permutations $S_3$ which, as
well-known, corresponds to the outer automorphisms (${O}ut\equiv
{A}ut/{I}nt$) of $D_4$.\par The groups $SO(8)$ and $SO(4,4)$ are
obtained by exponentiating different real forms of the $D_4$ Lie
algebra.\par For such groups and the corresponding metrics, the
euclidean one with either all $+$ or all $-$ signs and the
pseudoeuclidean metric $(++++----)$, the spinor representations
are Majorana-Weyl and satisfy the properties discussed in the
previous section. In particular chiral and antichiral spinors can
be consistently defined.\par A unique and ``miraculous" feature of
the above spacetimes, not shared by any other case, consists in
the fact that vectors, chiral and antichiral spinors (in short
V-C-A's) have the same dimensionality, being all
eight-components.\par Transformations exchanging V-C-A's are found
and, as we discuss later, they can all be identified. Such a
property can be visualized with the following triangle diagram
whose vertices are the interrelated V-C-A's:
\begin{eqnarray}
&\begin{array}{ccccc}
   &  & V &  &\\
   & \circ&  &\circ &\\
  C &  &\circ&  &A\\
\end{array}&
\label{triangle}
\end{eqnarray}
Vectors ($V_m$), chiral ($\psi_a$) and antichiral ($\chi_{\dot
a}$) spinors can be conveniently arranged into a single
$24$-component ``triality vector" $T$
\begin{eqnarray}
T_M&=& \left( \begin{array}{c}
  V_m \\
 {\psi_a} \\
{\chi_{\dot a}}
\end{array}\right)
\end{eqnarray}
whose Lorentz-transformation properties are given by
\begin{eqnarray}
T &\mapsto & T' = e^{\frac{1}{2}\omega_{mn} \Sigma^{mn}} \cdot T
\end{eqnarray}
where
\begin{eqnarray}
\Sigma^{mn} &=& {\Sigma_V}^{mn}\oplus {\Sigma_C}^{mn}\oplus
{\Sigma_A}^{mn}
\end{eqnarray}
while $({\Sigma_{V,C,A}})^{mn}$ are the Lorentz-generators for
vectors, chiral and antichiral spinors respectively:
\begin{eqnarray}
{({\Sigma_V}^{mn})_k}^l &=& {\delta^m}_k (g^{-1})^{nl} -
(g^{-1})^{ml} {\delta^n}_k\nonumber\\ {\Sigma_C}^{mn} &=&
\frac{1}{4}( \sigma^m {\tilde \sigma}^n -{\sigma^n }
{\tilde\sigma}^m)\nonumber\\ {\Sigma_A}^{mn} &=& \frac{1}{4}
({\tilde\sigma}^m { \sigma}^n -{\tilde\sigma^n } \sigma^m)
\end{eqnarray}
As far as the Lorentz transformation properties alone are
concerned, there is no need to discuss (and even introduce) the
character, commuting or anticommuting, of spinors. However, the
triality admits to be interpreted as an invariance property which
allows to introduce the ${\cal G}_{Tr}$ triality group of symmetry
(to be defined below). For such interpretation we need to specify
whether the spinors are assumed commuting or anticommuting.
Following Cartan we discuss here the case of commuting spinors. We
will comment the modifications occurring when anticommuting
spinors are taken into account\footnote{Commuting spinors too are
relevant in the physical literature, not only as supersymmetric
ghosts in a BRST quantization scheme; they also appear, e.g., in
the super-twistors quantization approach, see \cite{stwists}. The
fudamental reference on twistors is \cite{twists}.}.\par For both
values of $\eta =\pm 1$, the following bilinear Lorentz-invariants
can be introduced
\begin{eqnarray}
{\bf {\cal B}}_V &=& {V^T}_m ({g^{-1}})^{mn} V_n\nonumber\\
{\bf{\cal B}}_C &=& \psi^T C^{-1}\psi\nonumber \\
 {\bf{\cal B}}_A &=& \chi^T {\tilde C}^{-1}
\chi \label{bilinear}
\end{eqnarray}
(we use the conventions discussed in the previous section), while
the trilinear Lorentz-invariant
\begin{eqnarray}
{\bf{\cal T}} &=& \Psi^T {\cal C} \Gamma^m\Psi  \cdot V_m = 2
(\psi^T C^{-1} \sigma^m \chi \cdot V_m),\quad \quad \Psi \equiv
\left(
\begin{array}{c}
  \psi_a\\
  \chi_{\dot a}
\end{array}\right)
\label{triliter}
\end{eqnarray}
 is non-vanishing for $\eta =-1$. For anticommuting spinors the
 bilinears ${\bf{\cal B_C}}$, ${\bf {\cal B_A}}$ are identically
 vanishing, while ${\bf {\cal T}}$ is non-vanishing for $\eta
 =+1$.\par
 The group of invariances ${\cal G}$ is introduced as the group
 of linear homogeneous transformations acting on the $8\times
 3=24$ dimensional space of ``triality-vectors" leaving invariant,
 separately,
 ${\bf{\cal B_V}}$, ${\bf{\cal B_C}}$, ${\bf{\cal B_A}}$ and
${\bf{\cal T}}$.\par The group of triality ${\cal G}_{Tr}$ is
defined by relaxing one condition, as the group of
$24$-dimensional homogeneous linear transformations leaving
invariant ${\bf{\cal T}}$ and the total bilinear ${\bf{\cal
B}}_{Sum}$
\begin{eqnarray}
{\bf{\cal B}}_{Sum}&=& {\bf{\cal B_V}}+{\bf{\cal B_C}}+{\bf{\cal
B_A}}
\end{eqnarray}
It can be proven that ${\cal G}_{Tr}$ is given by the semidirect
product of ${\cal G}$ and the finite group $S_3$
\begin{eqnarray}
{\cal G}_{Tr} &=& {\cal G}\otimes_S S_3
\end{eqnarray}
 This result directly follows from the Cartan's
approach, even if it is not explicitly stated in the Cartan's
book\footnote{He introduced the analogs of the ${\cal P}$, ${\cal
R}$ transformations of formula (\ref{prvca}) which however, in his
case, do not satisfy the relations (\ref{pprr}) and cannot
therefore be taken as the generators of the $S_3$ group.}. The
transformations in $S_3$ are obtained from the generators ${\cal
P}$, ${\cal R}$ whose action, symbolically, is given by
\begin{eqnarray}
{\cal P} &:&\quad V \leftrightarrow V, \quad C\leftrightarrow A
\nonumber\\ {\cal R} &:& \quad V \leftrightarrow C, \quad
A\leftrightarrow A \label{prvca}
\end{eqnarray}
As a consequence ${\cal P}, {\cal R}$ can be decomposed into
eight-dimensional blocks matrices according to
\begin{eqnarray}
{\cal P} =\left(\begin{array}{ccc}
  P_1 & &  \\
   &  & P_2 \\
   & P_3 &
\end{array}\right),
\quad &\quad &\quad {\cal R} =\left(\begin{array}{ccc}
   &R_1 &  \\
  R_2 &  &  \\
   & &R_3
\end{array}\right)
\end{eqnarray}
${\cal P}$, ${\cal R}$ can be carefully chosen in such a way to
satisfy the relations
\begin{eqnarray}
{\cal P}^2= {\cal R}^2 = {\bf 1},\quad &\quad &\quad ({\cal
P}{\cal R})^3 = {\bf 1} \label{pprr}
\end{eqnarray}
showing their nature of $S_3$ generators.\par There is an
arbitrariness in the choice of ${\cal P}, {\cal R}$ since any
couple of gauge-transformed generators ${\hat {\cal P}} =
g\cdot{\cal P} \cdot g^{-1}$, ${\hat {\cal R}} = g\cdot{\cal R}
\cdot g^{-1}$, with $ g\in {\cal G}_{Tr}$, satisfy the same
properties and can be equally taken for generating $S_3$. It
should be mentioned that, under a Lorentz transformation realized
by $e^{\omega\cdot \Sigma}$, the generators ${\cal P}$ (${\cal
R}$) are mapped into ${\cal P}' = e^{\omega\cdot \Sigma}{\cal
P}{e^{\omega\cdot \Sigma}}^{-1}$ (and respectively ${\cal R}' =
e^{\omega\cdot \Sigma}{\cal R}{e^{\omega\cdot \Sigma}}^{-1}$).
However, since we are free to introduce a gauge-compensating
transformation, we can make invariant the choice of ${\cal P},
{\cal R}$ in any Lorentzian system of reference. As a consequence
the constants ${\cal P}, {\cal R}$ matrices can be introduced in
manifestly Lorentz-invariant actions if needed.
\par
It is not easy, due to the intertwined nature of their relations
(\ref{pprr}), to determine a concrete realization for ${\cal P},
{\cal R}$. We arrived at it with a trial-and-error procedure. The
final result is furnished in the appendices $2$-$4$ for each one
of the three possible metrics.\par If we disregard ${\bf{\cal
B}}_{Sum}$ and just demand the invariance of ${\bf{\cal T}}$, in
this case both ${\cal P}, {\cal R}$ can always be assumed
real-valued. This is no longer true when the invariance of
${\bf{\cal B}}_{Sum}$ is required, due to the fact that the three
metrics on V-C-A's can differ by an overall $-$ sign. This case is
easily managed by noticing that the transformations $P_2\mapsto
iP_2$, $P_3 \mapsto -i P_3$, $P_1$ unchanged (and $R_1\mapsto i
R_1$, $R_2\mapsto -i R_2$, $R_3$ unchanged) solves the problem
without altering neither the (\ref{pprr}) relations nor the
${\bf{\cal T}}$-invariance requirement.\par We finally comment
that vectors, chiral and antichiral spinors can be identified
under triality. If a basis in these three different spaces has
already be chosen, the most natural way of identifying them
($V\equiv X_1\cdot C$, $V\equiv Y_1\cdot A$) make use of the
eight-dimensional invertible operators ${X}_1$, ${Y}_1$ entering
\begin{eqnarray}
{\cal P}\cdot {\cal R} =\left(\begin{array}{ccc}
   & X_1&  \\
   &  & X_2 \\
  X_3 &  &
\end{array}\right),
\quad &\quad &\quad {\cal R}\cdot{\cal P}
=\left(\begin{array}{ccc}
   & &Y_1  \\
  Y_2 &  &  \\
   &Y_3 &
\end{array}\right)
\end{eqnarray}
The reason to use them is because $X_1$, $Y_1$ map into vectors
respectively chiral and antichiral spinors with transformations
which correspond to {\em even} permutations of V,C,A.

\vspace{0.2cm} \noindent{\section{The signature triality.}}

In this section we discuss other consequences of the $S_3$
automorphisms of $D_4$. Besides being responsible for the Cartan's
V-C-A triality in fact, triality properties are associated with
other structures. For purpose of clarity it will be convenient to
represent them symbolically with triangle diagrams as the one
shown in (\ref{triangle}).\par An extra-consequence of triality
appears at the level of Majorana-Weyl representations for
Clifford's $\Gamma$-matrices, see section $3$. The three different
eight-dimensional representations can in fact be placed into the
diagram
\begin{eqnarray}
&\begin{array}{ccccc}
   &  & (4_S+4_A) &  &\\
   & \circ&  &\circ &\\
  (8_S+0_A) &  &\circ&  &(0_S+8_A)\\
\end{array}&
\end{eqnarray}
which exhibits the triality operating at the level of the
$\Gamma$-matrices.\par We have recalled that such
MW-representations are associated with the space-time signature,
and therefore triality can also be regarded as operating on
space-times according to
\begin{eqnarray}
\left(\begin{array}{ccccc}
   &  & (5+5) &  &\\
   & \circ&  &\circ &\\
  (9+1) &  &\circ&  &(1+9)\\
\end{array}\right)
&\longrightarrow &\left(\begin{array}{ccccc}
   &  & (4+4) &  &\\
   & \circ&  &\circ &\\
  (8+0) &  &\circ&  &(0+8)\\
\end{array}\right)
\end{eqnarray}
The arrow has been inserted to recall that such triality can be
lifted to higher dimensions or, conversely, that the
$8$-dimensional spacetimes arise as transverse-coordinates spaces
in physical theories.\par The triality operating at the level of
spacetime signatures is the one visualized by the previous diagram
and will be described in this section.\par Cartan's V-C-A triality
and signature triality can also be combined and symbolically
represented by a sort of fractal-like double-triality diagram as
follows
\begin{eqnarray}
&\begin{array}{ccccc}
   &  &
\begin{array}{ccccc}
   &  & V &  &\\
   & \cdot&{\large{\bf (4+4)}}  &\cdot &\\
  C &  &\cdot&  &A\\
\end{array}
&  &\\ & &  & &\\ & &  & &\\ &\circ &  & \circ&\\ & &  & &\\ & &
& &\\
\begin{array}{ccccc}
&  & V &  &\\ & \cdot&{\large{\bf (8+0)}}  &\cdot &\\ C &  &\cdot&
&A\\
\end{array}
&  &\circ &  &
\begin{array}{ccccc}
&  & V &  &\\& \cdot&{\large{\bf (0+8)}}  &\cdot &\\ C &  &\cdot&
&A\\
\end{array}
\\
\end{array}&
\end{eqnarray}
The bigger triangle illustrates the signature triality, while the
smaller triangles visualize the trialities for vectors, chiral and
antichiral spinors living in each space-time.
\par
We give now the explicit expression of the duality transformations
relating theories formulated in the three above spacetimes or, in
other words, the ``translation rules" for the set of data
discussed in section $4$.
\par
To be definite we discuss the $\eta = -1$ case (we recall that the
$\eta$-sign has been introduced in section $2$); the modifications
to be introduced in the $\eta=+1$ case are immediate.\par Since we
are working in a Majorana-Weyl basis it is always true that ${\cal
B}= {\bf 1}$. The data defining our theories are therefore
specified by the metric in the vectors' space $g^{-1}$, as well as
the charge-conjugation matrix ${\cal C}$ which contains both the
metric for chiral ($C^{-1}$) and antichiral $({\tilde C}^{-1}$)
spinors. In a Majorana-Weyl basis the ${\cal A}$ matrix is
identified with ${\cal C}$ via the equation (\ref{1}).
\par
It is convenient to collectively denote as ${g_{\star}}^{-1}$
(where $\star\equiv V,C,A$) the three metrics associated to,
respectively, vectors and chiral and antichiral spinors in the
$(4+4)$ spacetime. The analogous metrics when associated to the
$(8+0)$ spacetime will be denoted with a tilde (${{\tilde
g}_{\star}}^{-1}$), while the hat will denote the three metrics
associated to the $(0+8)$ spacetime (${{\hat g}_{\star}}^{-1}$).
\par
Working in the $\eta=-1$ case with the representations given in
appendices $2-4$ we have
\begin{eqnarray}
&& {{g}_{V}}^{-1}= {\bf 1_4}\oplus {\bf - 1_4},\quad\quad {{
g}_{C}}^{-1}= {\bf 1_4}\oplus {\bf -1_4},\quad\quad {{
g}_{A}}^{-1}= {\bf 1_4}\oplus {\bf - 1_4};\nonumber\\ &&{{\tilde
g}_{V}}^{-1}= {\bf 1_8},\quad\quad {{\tilde g}_{C}}^{-1}= {\bf
1_8},\quad\quad {{\tilde g}_{A}}^{-1}= {\bf - 1_8};\nonumber\\
&&{{\hat g}_{V}}^{-1}= {\bf - 1_8},\quad\quad {{\hat g}_{C}}^{-1}=
{\bf 1_8},\quad\quad {{\hat g}_{A}}^{-1}= {\bf 1_8};
\end{eqnarray}
The duality transformations can therefore be expressed by
similarity transformations realized by non-orthogonal bridge
matrices connecting the different metrics. We can introduce indeed
the eight-dimensional bridge matrices ${\tilde K}_{\star}$, ${\hat
K}_{\star}$ such that
\begin{eqnarray}
{{\tilde g}_{\star}}^{-1} &=& {\tilde
K}_{\star}\cdot{g_{\star}}^{-1}\cdot {{\tilde
K}_{\star}}^T\nonumber\\ {{\hat g}_{\star}}^{-1} &=& {\hat
K}_{\star}\cdot{g_{\star}}^{-1} \cdot{{\hat K}_{\star}}^T
\end{eqnarray}
(as before ``$\star$" assumes the values $V$, $C$, $A$).\par
Vectors, chiral and antichiral spinors (V-C-A's, collectively
denoted as $\varphi_{\star}$) in the $(4+4)$ spacetime are
transformed  into the $(8+0)$-signature V-C-A's ${\tilde
\varphi}_{\star}$ according to
\begin{eqnarray}
{\tilde \varphi}_{\star} &\equiv & ({{\tilde K}_{\star}}^T)^{-1}
\cdot \varphi_{\star}
\end{eqnarray}
An analogous transformation maps them into the ($0+8)$-signature
V-C-A's ${\hat \varphi}_{\star}$.\par The $16$-dimensional
matrices ${\tilde H}$, ${\hat H}$ are constructed with the help of
${\tilde K}_C, {\tilde K}_A$ (and respectively ${\hat K}_C, {\hat
K}_A$) according to
\begin{eqnarray}
{\tilde H} &=& {\tilde K}_C \oplus {\tilde K}_A\nonumber\\ {\hat
H} &=& {\hat K}_C \oplus {\hat K}_A \end{eqnarray} They are used
to express the ($8+0$), and respectively ($0+8$),
charge-conjugation matrices ${\tilde {\cal C}}$ (${\hat {\cal
C}}$) in terms of the ($4+4$) charge-conjugation matrix ${\cal C}$
via the similarity transformation
\begin{eqnarray}
{\tilde {\cal C}} &=& {\tilde H} \cdot {\cal C} \cdot {\tilde H}^T
\end{eqnarray}
and the corresponding equation obtained by replacing
``${\tilde{\quad}}$" with ``${\hat{\quad}}$".\par For what
concerns the Clifford's $\Gamma$ matrices in the
($4+4$)-signature, they are mapped into the $(8+0)$ ${\tilde
\Gamma}$'s according to
\begin{eqnarray}
\Gamma^m &\mapsto & {\tilde \Gamma}^{\tilde m} = {({{\tilde
H}^{T}})^{-1}} \cdot \Gamma^m \cdot
 {\tilde H}^T {({{\tilde
K}_V}^T)_m}^{\tilde m}
\end{eqnarray}
As usual, an analogous relation maps them into the
$(0+8)$-signature ${\hat\Gamma}$'s.\par It is furthermore
convenient to introduce the set of $K_{\star}$ matrices, defined
through
\begin{eqnarray}
K_{\star} &=& {\tilde K}_{\star}\cdot {\hat K}_{\star},
\label{kkt}
\end{eqnarray}
connecting the $(8+0)$ with the $(0+8)$ signatures.\par
 Let us denote with ${\cal
W}$ (and respectively ${\cal Z}$) the transformations mapping the
different signatures set of data according to the following
symbolic actions:
\begin{eqnarray}
{\cal W} &:& (4+4)\leftrightarrow (4+4),\quad\quad (8+0)
\leftrightarrow (0+8)\nonumber\\ {\cal Z} &:&(4+4)\leftrightarrow
(8+0),\quad\quad (0+8) \leftrightarrow (0+8) \end{eqnarray} Such
transformations are explicitly realized by $24$-dimensional
matrices $W_{\star}$ (and respectively $Z_{\star}$) acting on
column vectors of the kind $(\phi_\star \quad {\tilde
\phi}_{\star} \quad {\hat \phi}_{\star} )$. They are expressed in
terms of the $8$-dimensional block matrices $K_\star$, ${\tilde
K}_{\star}$. Indeed we can put them into the form
\begin{eqnarray}
W_\star &=&\left(\begin{array}{ccc}
  {\bf 1_8} & 0 & 0 \\
  0 & 0 & K_{\star} \\
  0 & K_{\star} & 0
\end{array}\right)\nonumber\\
Z_\star &=&\left(\begin{array}{ccc}
  0 & {\tilde K}_\star & 0 \\
  {\tilde K}_\star& 0 & 0 \\
  0 & 0 & {\bf 1_8}
\end{array}\right)
\end{eqnarray}
The matrices ${\tilde K}_\star$, ${K}_\star$ can be carefully
chosen in such a way that the ${\cal W}$, ${\cal Z}$
transformations can be regarded as the generators of the $S_3$
group (i.e. the signature triality group). This implies of course
that the set of relations
\begin{eqnarray} {\cal W}^2 ={\cal Z}^2 &=& {\bf 1}\nonumber\\
({\cal W}\cdot{\cal Z})^3 &=& {\bf 1} \label{signtr}
\end{eqnarray}
must be satisfied.\par We mention here that when an odd number of
signatures is flipped, as it happens for the standard Wick
rotation from the Minkowski into the Euclidean space, the ${\bf
Z}_2$ group can not be realized with an action on the $\Gamma$'s
matrices. Indeed when, let's say, the $m=1$ direction is flipped,
the corresponding Clifford matrix $\Gamma^1$ is mapped into $
i\Gamma^1$, which leads to a ${\bf Z}_4$ group. A necessary
condition to realize a ${\bf Z}_2$ group on Clifford's $\Gamma$
matrices is that an {\em even} number of signatures has to be
flipped (as it happens when changing e.g. $(++)\mapsto (--)$).
This is due to the fact that the $\sigma_y= -i e_{12} + i e_{21}$
Pauli matrix, satisfying
\begin{eqnarray} {\sigma_y}^2 &=& {\bf 1}\end{eqnarray}
can be employed, allowing via a similarity transformation, to
switch the sign
\begin{eqnarray}
\sigma_y \cdot {\bf 1}_2 \cdot {\sigma_y}^T &=& - {\bf 1}_2
\end{eqnarray}
In the case here considered a consistent choice for $K_{\star}$,
${\tilde K}_{\star}$ is given by
\begin{eqnarray}
K_V &=&
\sigma_y\oplus\sigma_y\oplus\sigma_y\oplus\sigma_y\nonumber\\ K_C
&=& {\bf 1_8}\nonumber\\ K_A &=&
\sigma_y\oplus\sigma_y\oplus\sigma_y\oplus\sigma_y \nonumber\\
{\tilde K}_V &=& {\bf 1_4}\oplus \sigma_y\oplus
\sigma_y\nonumber\\ {\tilde K}_C &=& {\bf
1_4}\oplus\sigma_y\oplus\sigma_y\nonumber\\ {\tilde K}_A
&=&\sigma_y\oplus\sigma_y\oplus {\bf 1_4}
\end{eqnarray}
It is a straightforward exercise to verify the consistency of the
full set of relations (\ref{signtr}) with the above choice of
matrices. \par For what concerns ${\hat K}_\star$, they are
immediately read from (\ref{kkt}).\par The above construction
shows, as promised, that theories involving $8$-dimensional
Majorana-Weyl spinors in different signatures can all be dually
related in such a way to close the $S_3$ group of permutations.

\vspace{0.2cm} \noindent \section{An application: the
brane-molecule scan.}

We present in this section an application of the
signature-dualities properties induced by triality. It concerns
the so-called brane-molecule scan of ref. {\cite{Duff}},
originally appeared in the Blencowe-Duff paper in NPB, and later
revisited in the Duff's Tasi Lecture Notes. In these works the
conditions for the existence of classical supersymmetric branes of
arbitrary signatures embedded in flat target spaces whose
signatures too are left arbitrary, are analyzed in details. Such
conditions involve of course the property of spinors (Majorana,
Weyl or Majorana-Weyl), plus extra identities involving the
Clifford's $\Gamma$-matrices, which are needed in order to
implement the kappa-symmetry.\par The general result being
presented in \cite{Duff}, here we just recall the allowed branes
for $d=10,11,12$ dimensional targets. We have
\\
{\em i)} the superstrings $(1+1)\mapsto \{ (1+9), \quad(5+5)\quad
or \quad (9+1)\}$,\\ {\em ii)} the five-branes $\{(1+5)\mapsto \{
(1+9)\quad or\quad (5+5)\}$, as well as the mirror copies\par
$(5+1)\mapsto \{ (9+1)\quad or\quad (5+5)\}$, obtained by
exchanging space with time,
\\ {\em
iii)} the membranes $(1+2)\mapsto \{ (9+2),\quad (6+5),\quad
(5+6),\quad (2+9),\quad or \quad (1+10)\}$, as well as\par the
mirror copies obtained by space-time exchange,\\ {\em iv)}
finally, the non-minkowskian $(2+2) \mapsto \{ (2+10), \quad
(6+6)\quad or\quad (10+2)\}$ branes.
\par
In all the above cases the target space-times can be recovered
from the eight-dimensional spacetimes, according to the discussion
in section $3$. It is therefore clear that the ``translation
rules" expressing the various versions of the above theories in
different signatures can be expressed in terms of the previous
section results. In particular for the superstring, as well as for
the $(2+2)$ brane, the three different versions are dually mapped
in such a way to close the $S_3$ group.\par In the original papers
it was remarked the presence of a trivial ${\bf Z}_2$ symmetry
obtained by space-time exchange, connecting e.g. the $(1+9)$ to
the $(9+1)$ target spacetimes solutions of the superstring.
However, the presence of the extra solution $(5+5)$ was left
``unexplained". It turns out to be connected with, let's say, the
$(1+9)$ solution via another, non-trivial, generator of the $S_3$
triality group.\par The above triality property holds for other
classes of supersymmetric theories, like the $10$-dimensional
superYang-Mills theories, which are allowed in different
signatures (as before in the $(1+9), (5+5), (9+1)$ cases).

\vspace{0.2cm} \noindent \section{Triality as an invariance.}

In this section we wish to discuss another possible application of
triality related to the Duff's viewpoint that a fundamental theory
of everything should explain not only the dimensionality, but as
well the signature of the spacetime. According to the Hull and
Hull-Khuri results \cite{Hull}, the different versions in each
given signature of a given theory are dually mapped one into
another and therefore all equivalent. This is also the main
content of the previous sections analysis, where we pointed out
the role played by the $S_3$ triality group in such a context.\par
However our own results admit another possible interpretation.
Having identified $S_3$ as the duality group, a question that can
be naturally raised is whether this group can be assumed as a
group of invariance, providing a signature-independent framework
for the description of our theories. All this is much in the same
spirit as general relativity providing a coordinate independent
scheme.\par Such a question deserves of course a careful
investigation. It should be mentioned that it is not so difficult
to realize an $S_3$-triality invariant formulation for strings,
branes, etc. Other questions however, like the possibility that
the $S_3$ group could be spontaneously broken, have no answer at
present and need further investigation. A possible mechanism for
producing a spontaneous breaking could use potential terms induced
by the trilinear term given in formula (\ref{triliter}).\par The
most natural setting to investigate such questions seems to be the
eight-dimensional light-cone formulations of superstrings and
branes, since the triality can be manifestly realized in this
dimensionality.\par We conclude this section pointing out the
property that triality induced by supersymmetry strongly
constraints the number of finite groups allowing a
space-versus-time exchange symmetry. In bosonic theories, since
all signatures are consistent and therefore allowed, the number of
such groups is huge. In the supersymmetric case things are
different. Let us discuss the $10$-dimensional superstring
theories to be definite. The allowed signatures of the target
spacetimes are $(1+9), (5+5), (9+1)$. If a space-versus-time
coordinate exchange symmetry is required, as a consequence only
three groups arise, namely\\ {\em i)} the identity group ${\bf
1}$, corresponding to a theory formulated in the single spacetime
$(5+5)$,\\ {\em ii)} the ${\bf Z}_2$ group of symmetry, for a
theory formulated by using two spacetimes copies $(1+9)$,
\par $(9+1)$,
\\
{\em iii)} finally, the $S_3$ group, which underlines a unified
``space-time" theory requiring the whole \par set of
$10$-dimensional Majorana-Weyl spacetimes $(1+9), (5+5), (9+1)$.

\vspace{0.2cm} \noindent \section{Conclusions.}

In this paper we have investigated various consequences of the
triality automorphisms of $Spin(8)$. We have discussed in a
suitable Majorana-Weyl basis (and partially extended) the original
Cartan's results \cite{Cart} concerning the transformation
properties of $8$-dimensional vectors, chiral and antichiral
spinors. Moreover, we have shown how triality affects the
representation properties of the Clifford $\Gamma$ matrices (the
so-called Majorana-type representations). We have pointed out that
triality naturally encodes the property that supersymmetry in
higher-dimensions are consistently formulated in
different-signature spacetimes. The $S_3$ triality group quite
naturally arises in such a context. Indeed, we have been able to
prove that different signature-versions of a given supersymmetric
theory such, to give an example, as superstrings, can be dually
transformed one into another with trasformations, conveniently
chosen, which allow closing the $S_3$ group.\par The possible role
of $S_3$ as a symmetry group for a space-versus-time exchange
invariance has also been mentioned.\par One of the main
motivations to investigate such properties concerned the still
highly speculative but quite fascinating topics that a
supersymmetric fundamental theory could shed light on the nature
of the time. This point of view has been advocated by Duff
\cite{Duff} in a series of papers and recently gained increased
attention due to the results of Hull and collaborators
\cite{Hull}.
\par
The fact that supersymmetric theories seemingly related with the
supposed $F$-theory find a natural formulation in a two-time
physical world poses of course a strong challenge. Bars and
collaborators \cite{Bars}, e.g., are exploring the possibility
that the minkowksian time arises as a gauge-fixing of such a
two-time physical world.\par The present paper fits into this line
of research. It is worth stressing however that, no matter which
was the original motivation, the present paper contains a detailed
account of mathematical results and property of triality which can
be possibly used in technical analysis in different and more
down-to-earth contexts. Just to mention an example which is
currently under investigation, the here presented ``translation
rules" among different signatures can be used to define, let's
say, superYang-Mills theories in a $(5+5)$ signature. Such
theories can be dimensionally reduced to an AdS $(++---)$
signature. Standard minkowskian $10$-dimensional superYang-Mills
theories do not admit of course such a dimensional reduction to
AdS. \par It is furthermore worth mentioning that some of the
mathematical results here presented seems new, we have been unable
to find them in the existing mathematical literature, at least not
in such an explicit form as here given.

\vskip1cm \noindent{\Large{\bf Acknowledgments}} \\ {\quad}\\ We
are pleased to acknowledge J. A. Helay\"{e}l-Neto and L.P. Colatto
for both encouragement and helpful discussions. We are grateful to
DCP-CBPF for the kind hospitality. F.T. acknowledges financial
support from CNPq.

 \vskip1cm\noindent{\Large{\bf
Appendix $1$:}}\\ {\Large{\bf The $(0_S+6_A)$-representation for
the $6$-dimensional $\Gamma$'s.}}
\\{\quad}\par
For completeness we furnish here a representation of the
$6$-dimensional Clifford' s $\Gamma$-matrices realized in terms of
antisymmetric $\Gamma$'s only. As discussed in the text, this
realization allows to construct all Majorana-type representations
for Clifford's $\Gamma$-matrices up to dimension $d=12$ with the
help of the recursive formula (\ref{algo}). \par The
representation given below is the correct one to be used in order
to express a $6$-dimensional Euclidean space in the Majorana
basis, which is consistent for $t=6$, $s=0$ when $\eta = -1$, as
well as for $t=0$, $s=6$ and $\eta = +1$.\par We constructed the
$(0_S+6_A)$-representation out of the $(3_S+3_A)$-representation
(which is easily obtained through (\ref{algo})) after computing in
such a case, for one of the above Euclidean spaces, the value of
the ${\cal B}$-matrix and later finding a transformation $U$
mapping ${\cal B}\mapsto U^\ast {\cal B} U^\dagger = {\bf 1}$. As
a consequence we obtain ${\Gamma^i}_{(0S+6A)} = U
{\Gamma^i}_{(3S+3A)} U^\dagger $ according to (\ref{2}). The extra
$\Gamma^7$ matrix is also antisymmetric.\par The final result,
here presented for all $\Gamma$'s real (i.e. ${\Gamma^i}^2 =-{\bf
1}$), is given by:
\begin{eqnarray}
\Gamma^1 =\left(
\begin{array}{cccccccc}
  0 & 0 & 0 & 0 & 0 & 1 & 0 & 0 \\
  0 & 0 & 0 & 0 & -1 & 0 & 0 & 0 \\
  0 & 0 & 0 & 0 & 0 & 0 & 0 & -1 \\
  0 & 0 & 0 & 0 & 0 & 0 & 1 & 0 \\
  0 & 1 & 0 & 0 & 0 & 0 & 0 & 0 \\
  -1 & 0 & 0 & 0 & 0 & 0 & 0 & 0 \\
  0 & 0 & 0& -1& 0 & 0 & 0 & 0\\
  0& 0 & 1 & 0 & 0 & 0 & 0 & 0
\end{array}\right)\quad &&\quad
\Gamma^2 =\left(
\begin{array}{cccccccc}
  0 & 0 & 1 & 0 & 0 & 0 & 0 & 0 \\
  0 & 0 & 0 & -1 & 0 & 0 & 0 & 0 \\
  -1 & 0 & 0 & 0 & 0 & 0 & 0 & 0 \\
  0 & 1 & 0 & 0 & 0 & 0 & 0 & 0 \\
  0 & 0 & 0 & 0 & 0 & 0 & -1 & 0 \\
  0 & 0 & 0 & 0 & 0 & 0 & 0 & 1 \\
  0 & 0 & 0& 0 & 1 & 0 & 0 & 0\\
  0 & 0 & 0 & 0 & 0 & -1 & 0 & 0
\end{array}\right)\nonumber\\
\Gamma^3 = \left(
\begin{array}{cccccccc}
  0 & 0 & 0 & 1 & 0 & 0 & 0 & 0 \\
  0 & 0 & 1 & 0 & 0 & 0 & 0 & 0 \\
  0 & -1 & 0 & 0 & 0 & 0 & 0 & 0 \\
  -1 & 0 & 0 & 0 & 0 & 0 & 0 & 0 \\
  0 & 0 & 0 & 0 & 0 & 0 & 0 & -1 \\
  0 & 0 & 0 & 0 & 0 & 0 & -1 & 0 \\
  0 & 0 & 0& 0 & 0 & 1 & 0 & 0 \\
  0 & 0 & 0 & 0 & 1 & 0 & 0 & 0
\end{array}\right)
\quad&&\quad \Gamma^4 =\left(
\begin{array}{cccccccc}
  0 & 0 & 0 & 0 & 0 & 0 & -1 & 0 \\
  0 & 0 & 0 & 0 & 0 & 0 & 0 & -1 \\
  0 & 0 & 0 & 0 & 1 & 0 & 0 & 0 \\
  0 & 0 & 0 & 0 & 0 & 1 & 0 & 0 \\
  0 & 0 & -1 & 0 & 0 & 0 & 0 & 0 \\
  0 & 0 & 0 & -1 & 0 & 0 & 0 & 0 \\
  1 & 0 & 0 & 0 & 0 & 0 & 0 & 0 \\
  0 & 1 & 0 & 0 & 0 & 0 & 0 & 0
\end{array}\right)\nonumber\\
 \Gamma^5 =\left(
\begin{array}{cccccccc}
  0 & 1 & 0 & 0 & 0 & 0 & 0 & 0 \\
  -1 & 0 & 0 & 0 & 0 & 0 & 0 & 0 \\
  0 & 0 & 0 & 1 & 0 & 0 & 0 & 0 \\
  0 & 0 & -1 & 0 & 0 & 0 & 0 & 0 \\
  0 & 0 & 0 & 0 & 0 & -1 & 0 & 0 \\
  0 & 0 & 0 & 0 & 1 & 0 & 0 & 0 \\
  0 & 0 & 0& 0 & 0 & 0 & 0 & -1\\
  0 & 0 & 0 & 0 & 0 & 0 & 1 & 0
\end{array}\right)
\quad&&\quad \Gamma^6 = \left(
\begin{array}{cccccccc}
  0 & 0 & 0 & 0 & -1 & 0 & 0 & 0 \\
  0 & 0 & 0 & 0 & 0 & -1 & 0 & 0 \\
  0 & 0 & 0 & 0 & 0 & 0 & -1 & 0 \\
  0 & 0 & 0 & 0 & 0 & 0 & 0 & -1 \\
  1 & 0 & 0 & 0 & 0 & 0 & 0 & 0 \\
  0 & 1 & 0 & 0 & 0 & 0 & 0 & 0 \\
  0 & 0 & 1 & 0 & 0 & 0 & 0 & 0 \\
  0 & 0 & 0 & 1 & 0 & 0 & 0 & 0
\end{array}\right)\nonumber\\
\Gamma^7 = \left(
\begin{array}{cccccccc}
  0 & 0 & 0 & 0 & 0 & 0 & 0 & -1 \\
  0 & 0 & 0 & 0 & 0 & 0 & 1 & 0 \\
  0 & 0 & 0 & 0 & 0 & -1 & 0 & 0 \\
  0 & 0 & 0 & 0 & 1 & 0 & 0 & 0 \\
  0 & 0 & 0 & -1 & 0 & 0 & 0 & 0 \\
  0 & 0 & 1 & 0 & 0 & 0 & 0 & 0 \\
  0 & -1 & 0& 0 & 0 & 0 & 0 & 0 \\
  1 & 0 & 0 & 0 & 0 & 0 & 0 & 0
\end{array}\right)\quad&&\quad
\end{eqnarray}
\vskip1cm \noindent{\Large{\bf Appendix $2$:}}\\ {\quad}\\
{\Large{\bf The ($4_S+4_A$)-representation for the $8$-dimensional
$\Gamma$'s.}}\\ {\quad}
\par
In the following three appendices we explicitly furnish the three
Majorana-Weyl representations for the $8$-dimensional
$\Gamma$-matrices. Moreover in each case a specific realization of
the ${\cal P}$, ${\cal R}$ generators of the $S_3$ permutation
group, leaving invariant the trilinear term (\ref{triliter}) for
the choice $\eta = -1$, i.e. for {\em commuting} spinors, is given
(in the opposite case, $\eta=+1$, the trilinear term is
automatically vanishing for commuting spinors).\par The
$\Gamma$'s, as well as the generators ${\cal P},{\cal R}$, are
presented in real form (confront the discussion in section
$5$).\par Each one of the three representations, being MW, admits
a $\Gamma^9$ matrix of the kind
\begin{eqnarray}
\Gamma^9 &=& \left(\begin{array}{cc}
  {\bf 1_8} & 0 \\
  0 & {\bf - 1_8}
\end{array}\right)\nonumber
\end{eqnarray}
while the $\Gamma^i$ for $i=1,2,...,8$ are decomposed according to
\begin{eqnarray}
\Gamma^i &=& \left(\begin{array}{cc}
  0 & \sigma^i  \\
  {\tilde\sigma}^i & 0
\end{array}\right)
\label{blocks}
\end{eqnarray}
We point out that the $8$-dimensional $\Gamma$-matrices introduced
in this one and the two following appendices are not directly
obtainable through the recursion formula (\ref{algo}). Rather,
they have been conveniently chosen in order to provide a diagonal
metric for both chiral and antichiral spinors which coincides, up
to an overall sign, with the given diagonal metric for vectors.
\par In this appendix we present the results for the
($4_S+4_A$)-representation, the one which has to be used in order
to
introduce MW-spinors in a $t=4$, $s=4$ spacetime. \par We have
\begin{eqnarray} \sigma^1 =\left(
\begin{array}{cccccccc}
  0 & 0 & 0 & 0 & 0 & 0 & 0 & -1 \\
  0 & 0 & 0 & 0 & 0 & 0 & 1 & 0 \\
  0 & 0 & 0 & 0 & 0 & -1 & 0 & 0 \\
  0 & 0 & 0 & 0 & 1 & 0 & 0 & 0 \\
  0 & 0 & 0 & -1 & 0 & 0 & 0 & 0 \\
  0 & 0 & 1 & 0 & 0 & 0 & 0 & 0 \\
  0 & -1 & 0 & 0 & 0 & 0 & 0 & 0 \\
  1 & 0 & 0 & 0 & 0 & 0 & 0 & 0
\end{array}\right)&&
\sigma^2 =\left(
\begin{array}{cccccccc}
  0 & 0 & 0 & 0 & 0 & 0 & 1 & 0 \\
  0 & 0 & 0 & 0 & 0 & 0 & 0 & 1 \\
  0 & 0 & 0 & 0 & 1 & 0 & 0 & 0 \\
  0 & 0 & 0 & 0 & 0 & 1 & 0 & 0 \\
  0 & 0 & -1 & 0 & 0 & 0 & 0 & 0 \\
  0 & 0 & 0 & -1 & 0 & 0 & 0 & 0 \\
  1 & 0 & 0 & 0 & 0 & 0 & 0 & 0 \\
  0 & 1 & 0 & 0 & 0 & 0 & 0 & 0
\end{array}\right)\nonumber\\
\sigma^3 =\left(
\begin{array}{cccccccc}
  0 & 0 & 0 & 0 & 0 & -1 & 0& 0 \\
  0 & 0 & 0 & 0 & -1 & 0 & 0 & 0 \\
  0 & 0 & 0 & 0 & 0 & 0 & 0 & 1 \\
  0 & 0 & 0 & 0 & 0 & 0 & 1 & 0 \\
  0 & 1 & 0 & 0 & 0 & 0 & 0 & 0 \\
  -1 & 0 & 0 & 0 & 0 & 0 & 0 & 0 \\
  0 & 0 & 0 & -1 & 0 & 0 & 0 & 0 \\
  0 & 0 & 1 & 0 & 0 & 0 & 0 & 0
\end{array}\right)&&
\sigma^4 =\left(
\begin{array}{cccccccc}
  0 & 0 & 0 & 0 & 1 & 0 & 0 & 0 \\
  0 & 0 & 0 & 0 & 0 & -1 & 0 & 0 \\
  0 & 0 & 0 & 0 & 0 & 0 & -1 & 0 \\
  0 & 0 & 0 & 0 & 0 & 0 & 0 & 1 \\
  1 & 0 & 0 & 0 & 0 & 0 & 0 & 0 \\
  0 & 1 & 0 & 0 & 0 & 0 & 0 & 0 \\
  0 & 0 & 1 & 0 & 0 & 0 & 0 & 0 \\
  0 & 0 & 0 & 1 & 0 & 0 & 0 & 0
\end{array}\right)\nonumber\\
\sigma^5 =\left(
\begin{array}{cccccccc}
  -1 & 0 & 0 & 0 & 0 & 0 & 0 & 0 \\
  0 & 1 & 0 & 0 & 0 & 0 & 0 & 0 \\
  0 & 0 & 1 & 0 & 0 & 0 & 0 & 0 \\
  0 & 0 & 0 & 1 & 0 & 0 & 0 & 0 \\
  0 & 0 & 0 & 0 & -1 & 0 & 0 & 0 \\
  0 & 0 & 0 & 0 & 0 & -1 & 0 & 0 \\
  0 & 0 & 0 & 0 & 0 & 0 & -1 & 0 \\
  0 & 0 & 0 & 0 & 0 & 0 & 0 & 1
\end{array}\right)&&
\sigma^6 =\left(
\begin{array}{cccccccc}
  0 & -1 & 0 & 0 & 0 & 0 & 0 & 0 \\
  -1 & 0 & 0 & 0 & 0 & 0 & 0 & 0 \\
  0 & 0 & 0 & 1 & 0 & 0 & 0 & 0 \\
  0 & 0 & -1 & 0 & 0 & 0 & 0 & 0 \\
  0 & 0 & 0 & 0 & 0 & 1 & 0 & 0 \\
  0 & 0 & 0 & 0 & -1 & 0 & 0 & 0 \\
  0 & 0 & 0 & 0 & 0 & 0 & 0 & -1 \\
  0 & 0 & 0 & 0 & 0 & 0 & -1 & 0
\end{array}\right)\nonumber\\
\sigma^7 =\left(
\begin{array}{cccccccc}
  0 & 0 & -1 & 0 & 0 & 0 & 0 & 0 \\
  0 & 0 & 0 & -1 & 0 & 0 & 0 & 0 \\
  -1 & 0 & 0 & 0 & 0 & 0 & 0 & 0 \\
  0 & 1 & 0 & 0 & 0 & 0 & 0 & 0 \\
  0 & 0 & 0 & 0 & 0 & 0 & 1 & 0 \\
  0 & 0 & 0 & 0 & 0 & 0 & 0 & 1 \\
  0 & 0 & 0 & 0 & -1 & 0 & 0 & 0 \\
  0 & 0 & 0 & 0 & 0 & 1 & 0 & 0
\end{array}\right)&&
\sigma^8 =\left(
\begin{array}{cccccccc}
  0 & 0 & 0 & -1 & 0 & 0 & 0 & 0 \\
  0 & 0 & 1 & 0 & 0 & 0 & 0 & 0 \\
  0 & -1 & 0 & 0 & 0 & 0 & 0 & 0 \\
  -1 & 0 & 0 & 0 & 0 & 0 & 0 & 0 \\
  0 & 0 & 0 & 0 & 0 & 0 & 0 & -1 \\
  0 & 0 & 0 & 0 & 0 & 0 & 1 & 0 \\
  0 & 0 & 0 & 0 & 0 & -1 & 0 & 0 \\
  0 & 0 & 0 & 0 & -1 & 0 & 0 & 0
\end{array}\right)\nonumber
\end{eqnarray}
while ${\tilde\sigma}_i =-{\sigma_i}^T $ for $i=1, 2, 3,4$ and
${\tilde \sigma}_i ={\sigma_i}^T$ for $i=5, 6, 7,8$, (i.e. the
first four $\Gamma$-matrices are antisymmetric, the last four ones
symmetric).\par The diagonal charge-conjugation matrices are given
by
\begin{eqnarray}
C^{-1} &=& {\bf 1_4}\oplus {\bf - 1_4}\nonumber\\ {\tilde C}^{-1}
&=& - \eta \cdot C^{-1}
\end{eqnarray}
A convenient basis for ${\cal P}, {\cal R}$ is
\begin{eqnarray}
{\cal P}_{V\mapsto V} \equiv {\cal P}_1 &=&\left(
\begin{array}{cccccccc}
  -1 & 0 & 0 & 0 & 0 & 0 & 0 & 0 \\
  0 & -1 & 0 & 0 & 0 & 0 & 0 & 0 \\
  0 & 0 & -1 & 0 & 0 & 0 & 0 & 0 \\
  0 & 0 & 0 & -1 & 0 & 0 & 0 & 0 \\
  0 & 0 & 0 &  0 & -1 & 0 & 0 & 0 \\
  0 & 0 & 0 & 0 & 0 & -1 & 0 & 0 \\
  0 & 0 & 0 & 0 & 0 & 0 & -1 & 0 \\
  0 & 0 & 0 & 0 & 0 & 0 & 0 & 1
\end{array}\right)\nonumber\\
{\cal P}_{A\mapsto C} \equiv {\cal P}_2 &=& \left(
\begin{array}{cccccccc}
  0 & 0 & 0 & 1 & 0 & 0 & 0 & 0 \\
  0 & 0 & -1 & 0 & 0 & 0 & 0 & 0 \\
  0 & 1 & 0 & 0 & 0 & 0 & 0 & 0 \\
  1 & 0 & 0 & 0 & 0 & 0 & 0 & 0 \\
  0 & 0 & 0 & 0 & 0 & 0 & 0 & 1 \\
  0 & 0 & 0 & 0 & 0 & 0 & -1 & 0 \\
  0 & 0 & 0 & 0 & 0 & 1 & 0 & 0 \\
  0 & 0 & 0 & 0 & 1 & 0 & 0 & 0
\end{array}\right)\nonumber\\
 {\cal P}_{C\mapsto A} \equiv
{\cal P}_3 &=& \left(
\begin{array}{cccccccc}
  0 & 0 & 0 & 1 & 0 & 0 & 0 & 0 \\
  0 & 0 & 1 & 0 & 0 & 0 & 0 & 0 \\
  0 & -1 & 0 & 0 & 0 & 0 & 0 & 0 \\
  1 & 0 & 0 & 0 & 0 & 0 & 0 & 0 \\
  0 & 0 & 0 & 0 & 0 & 0 & 0 & 1 \\
  0 & 0 & 0 & 0 & 0 & 0 & 1 & 0 \\
  0 & 0 & 0 & 0 & 0 & -1 & 0 & 0 \\
  0 & 0 & 0 & 0 & 1 & 0 & 0 & 0
\end{array}\right)
\end{eqnarray}
and
\begin{eqnarray}
{\cal R}_{C\mapsto V} \equiv {\cal R}_1 &=&\left(
\begin{array}{cccccccc}
  0 & 0 & 0 & 0 & -1 & 0 & 0 & 0 \\
  0 & 0 & 0 & 0 & 0 & -1 & 0 & 0 \\
  0 & 0 & 0 & 0 & 0 & 0 & -1 & 0 \\
  0 & 0 & 0 & 0 & 0 & 0 & 0 & 1 \\
  0 & 0 & 0 & 1 & 0 & 0 & 0 & 0 \\
  0 & 0 & 1 & 0 & 0 & 0 & 0 & 0 \\
  0 & -1 & 0 & 0 & 0 & 0 & 0 & 0 \\
  -1 & 0 & 0 & 0 & 0 & 0 & 0 & 0
\end{array}\right)\nonumber\\
{\cal R}_{V\mapsto C} \equiv {\cal R}_2 &=& \left(
\begin{array}{cccccccc}
  0 & 0 & 0 & 0 & 0 & 0 & 0 & -1 \\
  0 & 0 & 0 & 0 & 0 & 0 & -1 & 0 \\
  0 & 0 & 0 & 0 & 0 & 1 & 0 & 0 \\
  0 & 0 & 0 & 0 & 1 & 0 & 0 & 0 \\
  -1 & 0 & 0 & 0 & 0 & 0 & 0 & 0 \\
  0 & -1 & 0 & 0 & 0 & 0 & 0 & 0 \\
  0 & 0 & -1 & 0 & 0 & 0 & 0 & 0 \\
  0 & 0 & 0 & 1 & 0 & 0 & 0 & 0
\end{array}\right)\nonumber\\
 {\cal R}_{A\mapsto A} \equiv
{\cal R}_3 &=& \left(
\begin{array}{cccccccc}
  -1 & 0 & 0 & 0 & 0 & 0 & 0 & 0 \\
  0 & -1 & 0 & 0 & 0 & 0 & 0 & 0 \\
  0 & 0 & -1 & 0 & 0 & 0 & 0 & 0 \\
  0 & 0 & 0 & 1 & 0 & 0 & 0 & 0 \\
  0 & 0 & 0 & 0 & -1 & 0 & 0 & 0 \\
  0 & 0 & 0 & 0 & 0 & -1 & 0 & 0 \\
  0 & 0 & 0 & 0 & 0 & 0 & -1 & 0 \\
  0 & 0 & 0 & 0 & 0 & 0 & 0 & -1
\end{array}\right)
\end{eqnarray}
\vskip1cm \noindent{\Large{\bf Appendix $3$:}}\\ {\quad}\\
{\Large{\bf The ($0_S+8_A$)-representation for the $8$-dimensional
$\Gamma$'s.}}\\ {\quad}
\par
This representation is the one to be used in order to introduce a
MW-basis for an Euclidean $t=8$, $s=0$ space when $\eta = -1$. In
this case all $\Gamma$'s have to be assumed imaginary (in the
following formulas we drop, as usual, the $i$).\par The following
formulas are directly obtainable from those presented in appendix
2 after applying a special case of the transformation (\ref{2})
discussed in section $2$. They are here presented for
completeness. We have
\begin{eqnarray}
\sigma^1 =\left(
\begin{array}{cccccccc}
  0 & 0 & 0 & 0 & 0 & 0 & 0 & -1 \\
  0 & 0 & 0 & 0 & 0 & 0 & 1 & 0 \\
  0 & 0 & 0 & 0 & 0 & -1 & 0 & 0 \\
  0 & 0 & 0 & 0 & 1 & 0 & 0 & 0 \\
  0 & 0 & 0 & -1 & 0 & 0 & 0 & 0 \\
  0 & 0 & 1 & 0 & 0 & 0 & 0 & 0 \\
  0 & -1 & 0 & 0 & 0 & 0 & 0 & 0 \\
  1 & 0 & 0 & 0 & 0 & 0 & 0 & 0
\end{array}\right)&&
\sigma^2 =\left(
\begin{array}{cccccccc}
  0 & 0 & 0 & 0 & 0 & 0 & 1 & 0 \\
  0 & 0 & 0 & 0 & 0 & 0 & 0 & 1 \\
  0 & 0 & 0 & 0 & 1 & 0 & 0 & 0 \\
  0 & 0 & 0 & 0 & 0 & 1 & 0 & 0 \\
  0 & 0 & -1 & 0 & 0 & 0 & 0 & 0 \\
  0 & 0 & 0 & -1 & 0 & 0 & 0 & 0 \\
  1 & 0 & 0 & 0 & 0 & 0 & 0 & 0 \\
  0 & 1 & 0 & 0 & 0 & 0 & 0 & 0
\end{array}\right)\nonumber\\
\sigma^3 =\left(
\begin{array}{cccccccc}
  0 & 0 & 0 & 0 & 0 & -1 & 0& 0 \\
  0 & 0 & 0 & 0 & -1 & 0 & 0 & 0 \\
  0 & 0 & 0 & 0 & 0 & 0 & 0 & 1 \\
  0 & 0 & 0 & 0 & 0 & 0 & 1 & 0 \\
  0 & 1 & 0 & 0 & 0 & 0 & 0 & 0 \\
  -1 & 0 & 0 & 0 & 0 & 0 & 0 & 0 \\
  0 & 0 & 0 & -1 & 0 & 0 & 0 & 0 \\
  0 & 0 & 1 & 0 & 0 & 0 & 0 & 0
\end{array}\right)&&
\sigma^4 =\left(
\begin{array}{cccccccc}
  0 & 0 & 0 & 0 & 1 & 0 & 0 & 0 \\
  0 & 0 & 0 & 0 & 0 & -1 & 0 & 0 \\
  0 & 0 & 0 & 0 & 0 & 0 & -1 & 0 \\
  0 & 0 & 0 & 0 & 0 & 0 & 0 & 1 \\
  1 & 0 & 0 & 0 & 0 & 0 & 0 & 0 \\
  0 & 1 & 0 & 0 & 0 & 0 & 0 & 0 \\
  0 & 0 & 1 & 0 & 0 & 0 & 0 & 0 \\
  0 & 0 & 0 & 1 & 0 & 0 & 0 & 0
\end{array}\right)\nonumber\\
\sigma^5 =\left(
\begin{array}{cccccccc}
  1 & 0 & 0 & 0 & 0 & 0 & 0 & 0 \\
  0 & -1 & 0 & 0 & 0 & 0 & 0 & 0 \\
  0 & 0 & -1 & 0 & 0 & 0 & 0 & 0 \\
  0 & 0 & 0 & -1 & 0 & 0 & 0 & 0 \\
  0 & 0 & 0 & 0 & -1 & 0 & 0 & 0 \\
  0 & 0 & 0 & 0 & 0 & -1 & 0 & 0 \\
  0 & 0 & 0 & 0 & 0 & 0 & -1 & 0 \\
  0 & 0 & 0 & 0 & 0 & 0 & 0 & 1
\end{array}\right)&&
\sigma^6 =\left(
\begin{array}{cccccccc}
  0 & 1 & 0 & 0 & 0 & 0 & 0 & 0 \\
  1 & 0 & 0 & 0 & 0 & 0 & 0 & 0 \\
  0 & 0 & 0 & -1 & 0 & 0 & 0 & 0 \\
  0 & 0 & 1 & 0 & 0 & 0 & 0 & 0 \\
  0 & 0 & 0 & 0 & 0 & 1 & 0 & 0 \\
  0 & 0 & 0 & 0 & -1 & 0 & 0 & 0 \\
  0 & 0 & 0 & 0 & 0 & 0 & 0 & -1 \\
  0 & 0 & 0 & 0 & 0 & 0 & -1 & 0
\end{array}\right)\nonumber\\
\sigma^7 =\left(
\begin{array}{cccccccc}
  0 & 0 & 1 & 0 & 0 & 0 & 0 & 0 \\
  0 & 0 & 0 & 1 & 0 & 0 & 0 & 0 \\
  1 & 0 & 0 & 0 & 0 & 0 & 0 & 0 \\
  0 & -1 & 0 & 0 & 0 & 0 & 0 & 0 \\
  0 & 0 & 0 & 0 & 0 & 0 & 1 & 0 \\
  0 & 0 & 0 & 0 & 0 & 0 & 0 & 1 \\
  0 & 0 & 0 & 0 & -1 & 0 & 0 & 0 \\
  0 & 0 & 0 & 0 & 0 & 1 & 0 & 0
\end{array}\right)&&
\sigma^8 =\left(
\begin{array}{cccccccc}
  0 & 0 & 0 & 1 & 0 & 0 & 0 & 0 \\
  0 & 0 & -1 & 0 & 0 & 0 & 0 & 0 \\
  0 & 1 & 0 & 0 & 0 & 0 & 0 & 0 \\
  1 & 0 & 0 & 0 & 0 & 0 & 0 & 0 \\
  0 & 0 & 0 & 0 & 0 & 0 & 0 & -1 \\
  0 & 0 & 0 & 0 & 0 & 0 & 1 & 0 \\
  0 & 0 & 0 & 0 & 0 & -1 & 0 & 0 \\
  0 & 0 & 0 & 0 & -1 & 0 & 0 & 0
\end{array}\right)\nonumber
\end{eqnarray}
Since the $\Gamma^i$ are antisymmetric it follows that
${\tilde\sigma}^i =-{\sigma^i}^T $ for any $i=1,2,...,8$.\par The
two diagonal charge-conjugation matrices are
\begin{eqnarray}
C^{-1} &=& {\bf 1_8}\nonumber\\ {\tilde C}^{-1} &=& \eta \cdot
{\bf 1_8}
\end{eqnarray}
A real basis for ${\cal P}, {\cal R}$ is obtained by transforming
the corresponding mappings in the $(4_S+4_A)$ case. It is given by
\begin{eqnarray}
{\cal P}_{V\mapsto V} \equiv {\cal P}_1 &=&\left(
\begin{array}{cccccccc}
  -1 & 0 & 0 & 0 & 0 & 0 & 0 & 0 \\
  0 & -1 & 0 & 0 & 0 & 0 & 0 & 0 \\
  0 & 0 & -1 & 0 & 0 & 0 & 0 & 0 \\
  0 & 0 & 0 & -1 & 0 & 0 & 0 & 0 \\
  0 & 0 & 0 &  0 & -1 & 0 & 0 & 0 \\
  0 & 0 & 0 & 0 & 0 & -1 & 0 & 0 \\
  0 & 0 & 0 & 0 & 0 & 0 & 1 & 0 \\
  0 & 0 & 0 & 0 & 0 & 0 & 0 & -1
\end{array}\right)\nonumber\\
{\cal P}_{A\mapsto C} \equiv {\cal P}_2 &=& \left(
\begin{array}{cccccccc}
  0 & 0 & 1 & 0 & 0 & 0 & 0 & 0 \\
  0 & 0 & 0 & 1 & 0 & 0 & 0 & 0 \\
  1 & 0 & 0 & 0 & 0 & 0 & 0 & 0 \\
  0 & -1 & 0 & 0 & 0 & 0 & 0 & 0 \\
  0 & 0 & 0 & 0 & 0 & 0 & 1 & 0 \\
  0 & 0 & 0 & 0 & 0 & 0 & 0 & 1 \\
  0 & 0 & 0 & 0 & -1 & 0 & 0 & 0 \\
  0 & 0 & 0 & 0 & 0 & 1 & 0 & 0
\end{array}\right)\nonumber\\
 {\cal P}_{C\mapsto A} \equiv
{\cal P}_3 &=& \left(
\begin{array}{cccccccc}
  0 & 0 & 1 & 0 & 0 & 0 & 0 & 0 \\
  0 & 0 & 0 & -1 & 0 & 0 & 0 & 0 \\
  1 & 0 & 0 & 0 & 0 & 0 & 0 & 0 \\
  0 & 1 & 0 & 0 & 0 & 0 & 0 & 0 \\
  0 & 0 & 0 & 0 & 0 & 0 & -1 & 0 \\
  0 & 0 & 0 & 0 & 0 & 0 & 0 & 1 \\
  0 & 0 & 0 & 0 & 1 & 0 & 0 & 0 \\
  0 & 0 & 0 & 0 & 0 & 1 & 0 & 0
\end{array}\right)
\end{eqnarray}
and
\begin{eqnarray}
{\cal R}_{C\mapsto V} \equiv {\cal R}_1 &=&\left(
\begin{array}{cccccccc}
  0 & 0 & 0 & 0 & 0 & 1 & 0 & 0 \\
  0 & 0 & 0 & 0 & -1 & 0 & 0 & 0 \\
  0 & 0 & 0 & 0 & 0 & 0 & 0 & 1 \\
  0 & 0 & 0 & 0 & 0 & 0 & 1 & 0 \\
  0 & 0 & -1 & 0 & 0 & 0 & 0 & 0 \\
  0 & 0 & 0 & 1 & 0 & 0 & 0 & 0 \\
  1& 0 & 0 & 0 & 0 & 0 & 0 & 0 \\
  0 & -1 & 0 & 0 & 0 & 0 & 0 & 0
\end{array}\right)\nonumber\\
{\cal R}_{V\mapsto C} \equiv {\cal R}_2 &=& \left(
\begin{array}{cccccccc}
  0 & 0 & 0 & 0 & 0 & 0 & 1 & 0 \\
  0 & 0 & 0 & 0 & 0 & 0 & 0 & -1 \\
  0 & 0 & 0 & 0 & -1 & 0 & 0 & 0 \\
  0 & 0 & 0 & 0 & 0 & 1 & 0 & 0 \\
  0 & -1 & 0 & 0 & 0 & 0 & 0 & 0 \\
  1 & 0 & 0 & 0 & 0 & 0 & 0 & 0 \\
  0 & 0 & 0 & 1 & 0 & 0 & 0 & 0 \\
  0 & 0 & 1 & 0 & 0 & 0 & 0 & 0
\end{array}\right)\nonumber\\
 {\cal R}_{A\mapsto A} \equiv
{\cal R}_3 &=& \left(
\begin{array}{cccccccc}
  -1 & 0 & 0 & 0 & 0 & 0 & 0 & 0 \\
  0 & -1 & 0 & 0 & 0 & 0 & 0 & 0 \\
  0 & 0 & 1 & 0 & 0 & 0 & 0 & 0 \\
  0 & 0 & 0 & -1 & 0 & 0 & 0 & 0 \\
  0 & 0 & 0 & 0 & -1 & 0 & 0 & 0 \\
  0 & 0 & 0 & 0 & 0 & -1 & 0 & 0 \\
  0 & 0 & 0 & 0 & 0 & 0 & -1 & 0 \\
  0 & 0 & 0 & 0 & 0 & 0 & 0 & -1
\end{array}\right)
\end{eqnarray}
\vskip1cm \noindent{\Large{\bf Appendix $4$:}}\\ {\quad}\\
{\Large{\bf The ($8_S+0_A$)-representation for the $8$-dimensional
$\Gamma$'s.}}\\ {\quad}
\par
This representation is the one to be used in order to introduce a
MW-basis for an Euclidean $t=0$, $s=8$ space when $\eta = -1$. In
this case as well as all $\Gamma$'s have to be assumed imaginary
(in the following formulas we drop, as usual, the $i$).\par As
before the following formulas are obtained after transforming the
corresponding formulas in the ($4_S+4_A$) case. We have
\begin{eqnarray}
\sigma^1 =\left(
\begin{array}{cccccccc}
  0 & 0 & 0 & 0 & 0 & 0 & 0 & 1 \\
  0 & 0 & 0 & 0 & 0 & 0 & -1 & 0 \\
  0 & 0 & 0 & 0 & 0 & 1 & 0 & 0 \\
  0 & 0 & 0 & 0 & -1 & 0 & 0 & 0 \\
  0 & 0 & 0 & -1 & 0 & 0 & 0 & 0 \\
  0 & 0 & 1 & 0 & 0 & 0 & 0 & 0 \\
  0 & 1 & 0 & 0 & 0 & 0 & 0 & 0 \\
  -1 & 0 & 0 & 0 & 0 & 0 & 0 & 0
\end{array}\right)&&
\sigma^2 =\left(
\begin{array}{cccccccc}
  0 & 0 & 0 & 0 & 0 & 0 & -1 & 0 \\
  0 & 0 & 0 & 0 & 0 & 0 & 0 & -1 \\
  0 & 0 & 0 & 0 & -1 & 0 & 0 & 0 \\
  0 & 0 & 0 & 0 & 0 & -1 & 0 & 0 \\
  0 & 0 & -1 & 0 & 0 & 0 & 0 & 0 \\
  0 & 0 & 0 & -1 & 0 & 0 & 0 & 0 \\
  -1 & 0 & 0 & 0 & 0 & 0 & 0 & 0 \\
  0 & -1 & 0 & 0 & 0 & 0 & 0 & 0
\end{array}\right)\nonumber\\
\sigma^3 =\left(
\begin{array}{cccccccc}
  0 & 0 & 0 & 0 & 0 & 1 & 0& 0 \\
  0 & 0 & 0 & 0 & 1 & 0 & 0 & 0 \\
  0 & 0 & 0 & 0 & 0 & 0 & 0 & -1 \\
  0 & 0 & 0 & 0 & 0 & 0 & -1 & 0 \\
  0 & 1 & 0 & 0 & 0 & 0 & 0 & 0 \\
  -1 & 0 & 0 & 0 & 0 & 0 & 0 & 0 \\
  0 & 0 & 0 & 1 & 0 & 0 & 0 & 0 \\
  0 & 0 & -1 & 0 & 0 & 0 & 0 & 0
\end{array}\right)&&
\sigma^4 =\left(
\begin{array}{cccccccc}
  0 & 0 & 0 & 0 & -1 & 0 & 0 & 0 \\
  0 & 0 & 0 & 0 & 0 & 1 & 0 & 0 \\
  0 & 0 & 0 & 0 & 0 & 0 & 1 & 0 \\
  0 & 0 & 0 & 0 & 0 & 0 & 0 & -1 \\
  1 & 0 & 0 & 0 & 0 & 0 & 0 & 0 \\
  0 & 1 & 0 & 0 & 0 & 0 & 0 & 0 \\
  0 & 0 & -1 & 0 & 0 & 0 & 0 & 0 \\
  0 & 0 & 0 & -1 & 0 & 0 & 0 & 0
\end{array}\right)\nonumber\\
\sigma^5 =\left(
\begin{array}{cccccccc}
  -1 & 0 & 0 & 0 & 0 & 0 & 0 & 0 \\
  0 & 1 & 0 & 0 & 0 & 0 & 0 & 0 \\
  0 & 0 & 1 & 0 & 0 & 0 & 0 & 0 \\
  0 & 0 & 0 & 1 & 0 & 0 & 0 & 0 \\
  0 & 0 & 0 & 0 & -1 & 0 & 0 & 0 \\
  0 & 0 & 0 & 0 & 0 & -1 & 0 & 0 \\
  0 & 0 & 0 & 0 & 0 & 0 & 1 & 0 \\
  0 & 0 & 0 & 0 & 0 & 0 & 0 & -1
\end{array}\right)&&
\sigma^6 =\left(
\begin{array}{cccccccc}
  0 & -1 & 0 & 0 & 0 & 0 & 0 & 0 \\
  -1 & 0 & 0 & 0 & 0 & 0 & 0 & 0 \\
  0 & 0 & 0 & 1 & 0 & 0 & 0 & 0 \\
  0 & 0 & -1 & 0 & 0 & 0 & 0 & 0 \\
  0 & 0 & 0 & 0 & 0 & 1 & 0 & 0 \\
  0 & 0 & 0 & 0 & -1 & 0 & 0 & 0 \\
  0 & 0 & 0 & 0 & 0 & 0 & 0 & 1 \\
  0 & 0 & 0 & 0 & 0 & 0 & 1 & 0
\end{array}\right)\nonumber\\
\sigma^7 =\left(
\begin{array}{cccccccc}
  0 & 0 & -1 & 0 & 0 & 0 & 0 & 0 \\
  0 & 0 & 0 & -1 & 0 & 0 & 0 & 0 \\
  -1 & 0 & 0 & 0 & 0 & 0 & 0 & 0 \\
  0 & 1 & 0 & 0 & 0 & 0 & 0 & 0 \\
  0 & 0 & 0 & 0 & 0 & 0 & 1 & 0 \\
  0 & 0 & 0 & 0 & 0 & 0 & 0 & 1 \\
  0 & 0 & 0 & 0 & 1 & 0 & 0 & 0 \\
  0 & 0 & 0 & 0 & 0 & -1 & 0 & 0
\end{array}\right)&&
\sigma^8 =\left(
\begin{array}{cccccccc}
  0 & 0 & 0 & -1 & 0 & 0 & 0 & 0 \\
  0 & 0 & 1 & 0 & 0 & 0 & 0 & 0 \\
  0 & -1 & 0 & 0 & 0 & 0 & 0 & 0 \\
  -1 & 0 & 0 & 0 & 0 & 0 & 0 & 0 \\
  0 & 0 & 0 & 0 & 0 & 0 & 0 & -1 \\
  0 & 0 & 0 & 0 & 0 & 0 & 1 & 0 \\
  0 & 0 & 0 & 0 & 0 & 1 & 0 & 0 \\
  0 & 0 & 0 & 0 & 1 & 0 & 0 & 0
\end{array}\right)\nonumber
\end{eqnarray}
while now ${\tilde\sigma}^i ={\sigma^i}^T $ for any
$i=1,2,...,8$.\par The two diagonal charge conjugation matrices
are given by
\begin{eqnarray}
C^{-1} &=& {\bf 1_8}\nonumber\\ {\tilde C}^{-1} &=& -\eta\cdot{\bf
1_8}
\end{eqnarray} A real basis for ${\cal P}, {\cal R}$ is given by
\begin{eqnarray}
{\cal P}_{V\mapsto V} \equiv {\cal P}_1 &=&\left(
\begin{array}{cccccccc}
  -1 & 0 & 0 & 0 & 0 & 0 & 0 & 0 \\
  0 & -1 & 0 & 0 & 0 & 0 & 0 & 0 \\
  0 & 0 & -1 & 0 & 0 & 0 & 0 & 0 \\
  0 & 0 & 0 & -1 & 0 & 0 & 0 & 0 \\
  0 & 0 & 0 &  0 & -1 & 0 & 0 & 0 \\
  0 & 0 & 0 & 0 & 0 & -1 & 0 & 0 \\
  0 & 0 & 0 & 0 & 0 & 0 & -1 & 0 \\
  0 & 0 & 0 & 0 & 0 & 0 & 0 & 1
\end{array}\right)\nonumber\\
{\cal P}_{A\mapsto C} \equiv {\cal P}_2 &=& \left(
\begin{array}{cccccccc}
  0 & 0 & 0 & 1 & 0 & 0 & 0 & 0 \\
  0 & 0 & -1 & 0 & 0 & 0 & 0 & 0 \\
  0 & 1 & 0 & 0 & 0 & 0 & 0 & 0 \\
  1 & 0 & 0 & 0 & 0 & 0 & 0 & 0 \\
  0 & 0 & 0 & 0 & 0 & 0 & 0 & 1 \\
  0 & 0 & 0 & 0 & 0 & 0 & -1 & 0 \\
  0 & 0 & 0 & 0 & 0 & -1 & 0 & 0 \\
  0 & 0 & 0 & 0 & 1 & 0 & 0 & 0
\end{array}\right)\nonumber\\
 {\cal P}_{C\mapsto A} \equiv
{\cal P}_3 &=& \left(
\begin{array}{cccccccc}
  0 & 0 & 0 & 1 & 0 & 0 & 0 & 0 \\
  0 & 0 & 1 & 0 & 0 & 0 & 0 & 0 \\
  0 & -1 & 0 & 0 & 0 & 0 & 0 & 0 \\
  1 & 0 & 0 & 0 & 0 & 0 & 0 & 0 \\
  0 & 0 & 0 & 0 & 0 & 0 & 0 & -1 \\
  0 & 0 & 0 & 0 & 0 & 0 & -1 & 0 \\
  0 & 0 & 0 & 0 & 0 & -1 & 0 & 0 \\
  0 & 0 & 0 & 0 & 1 & 0 & 0 & 0
\end{array}\right)
\end{eqnarray}
and
\begin{eqnarray}
{\cal R}_{C\mapsto V} \equiv {\cal R}_1 &=&\left(
\begin{array}{cccccccc}
  0 & 0 & 0 & 0 & -1 & 0 & 0 & 0 \\
  0 & 0 & 0 & 0 & 0 & -1 & 0 & 0 \\
  0 & 0 & 0 & 0 & 0 & 0 & 1 & 0 \\
  0 & 0 & 0 & 0 & 0 & 0 & 0 & -1 \\
  0 & 0 & 0 & 1 & 0 & 0 & 0 & 0 \\
  0 & 0 & 1 & 0 & 0 & 0 & 0 & 0 \\
  0 & -1 & 0 & 0 & 0 & 0 & 0 & 0 \\
  -1 & 0 & 0 & 0 & 0 & 0 & 0 & 0
\end{array}\right)\nonumber\\
{\cal R}_{V\mapsto C} \equiv {\cal R}_2 &=& \left(
\begin{array}{cccccccc}
  0 & 0 & 0 & 0 & 0 & 0 & 0 & -1 \\
  0 & 0 & 0 & 0 & 0 & 0 & -1 & 0 \\
  0 & 0 & 0 & 0 & 0 & 1 & 0 & 0 \\
  0 & 0 & 0 & 0 & 1& 0 & 0 & 0 \\
  -1 & 0 & 0 & 0 & 0 & 0 & 0 & 0 \\
  0 & -1 & 0 & 0 & 0 & 0 & 0 & 0 \\
  0 & 0 & 1 & 0 & 0 & 0 & 0 & 0 \\
  0 & 0 & 0 & -1 & 0 & 0 & 0 & 0
\end{array}\right)\nonumber\\
 {\cal R}_{A\mapsto A} \equiv
{\cal R}_3 &=& \left(
\begin{array}{cccccccc}
  -1 & 0 & 0 & 0 & 0 & 0 & 0 & 0 \\
  0 & -1 & 0 & 0 & 0 & 0 & 0 & 0 \\
  0 & 0 & -1 & 0 & 0 & 0 & 0 & 0 \\
  0 & 0 & 0 & 1 & 0 & 0 & 0 & 0 \\
  0 & 0 & 0 & 0 & -1 & 0 & 0 & 0 \\
  0 & 0 & 0 & 0 & 0 & -1 & 0 & 0 \\
  0 & 0 & 0 & 0 & 0 & 0 & -1 & 0 \\
  0 & 0 & 0 & 0 & 0 & 0 & 0 & -1
\end{array}\right)
\end{eqnarray}

\end{document}